\documentclass[12pt,onecolumn,aps,pra,floatfix,superscriptaddress,preprint, showpacs,groupedaddress]{revtex4-1}

\usepackage{graphicx}
\usepackage{graphics}
\usepackage{amssymb}
\usepackage{amsmath}
\usepackage{epsfig}
\usepackage{latexsym}
\usepackage{color}
\usepackage{rotating}
\usepackage{subfigure}
\usepackage{float}
\usepackage[breaklinks=true]{hyperref}

\hypersetup{
	colorlinks   = true, 
	urlcolor     = blue, 
	linkcolor    = blue, 
	citecolor   =  red 
}

\usepackage{soul} 

\begin{document}
	
	\title{Precision ultrasound sensing on a chip}

	\author{Sahar Basiri-Esfahani$^{1,2}$,
		Ardalan Armin$^{1,2}$,
		Stefan Forstner$^1$,
		and Warwick P. Bowen$^1$\footnote{w.bowen@uq.edu.au}}
	\affiliation{$^1$ARC Centre for Engineered Quantum Systems, School of Mathematics and Physics, The University of Queensland, St Lucia, QLD 4072, Australia}
	\affiliation{$^2$Department of Physics, Swansea University, Singleton Park, Swansea SA2 8PP, Wales, United Kingdom }

\begin{abstract}

Ultrasound sensors have wide applications across science and technology. However, improved sensitivity is required for both miniaturisation and increased spatial resolution. Here, we introduce cavity optomechanical ultrasound sensing, where dual optical and mechanical resonances enhance the ultrasound signal. We achieve noise equivalent pressures of  8--300~$\mu$Pa/$\sqrt{\rm Hz}$ at kilohertz to megahertz frequencies in a microscale silicon-chip-based sensor with $>$120 dB dynamic range. The  sensitivity far exceeds similar sensors that use optical resonance alone and, normalised to sensing area, surpasses previous air-coupled ultrasound sensors by several orders of magnitude. The noise floor is, for the first time, dominated by collisions from molecules in the gas within which the acoustic wave propagates. This new approach to acoustic sensing could find applications ranging from biomedical diagnostics, to autonomous navigation, trace gas sensing, and scientific exploration of the life-induced-vibrations of single cells.

\end{abstract}

\maketitle


Over the past decade, cavity optomechanical sensors have emerged as a new class of ultraprecise photonic sensors~\cite{Purdy2017,KippenbergForce,PainterAccelerometer,WBmagnetometer,Basiri-optical-sensor}. These sensors integrate a high quality mechanical resonator with a high quality optical cavity (see e.g.~\cite{PainterPhC}). The mechanical resonator amplifies the mechanical vibrations introduced by resonant signals and provides isolation from environmental thermal noise, while the cavity resonantly enhances the optical response to the mechanical vibrations. A characteristic feature of cavity optomechanical sensors is that they are often only limited by optical shot noise and mechanical thermal noise, allowing the intrinsic limits in sensing performance to be approached~\cite{BowenMilburn}. This provides the ability to perform exquisitely sensitive optical measurements, with sub-attometre precision~\cite{Schliesser2008}. At kilometre scales it has proved crucial for the successful
detection of gravitational waves~\cite{gravit wave}; while at micro- and nano-scales it has enabled high performance acceleration, single-molecule, temperature and magnetic field sensing~\cite{Purdy2017, PainterAccelerometer, Lu2016, Wu, WBmagnetometer,WBmagnetometer2, Kim2017}, as well as
provided a new approach to control the quantum physics of massive objects, allowing quantum ground-state cooling~\cite{Teufel,Chan,Schliesser} and the generation of macroscopic non-classical states of motion~\cite{groblacher, Schwab}, with applications in future quantum technologies~(for e.g. see \cite{Andrews2014,Chu2017,Basiri-Esfahani}).

Detection of acoustic waves is essential for many applications including medical diagnostics, sonar, navigation, trace gas sensing and industrial processes~\cite{Dong,Fischer}. Most acoustic sensors transform an acoustic pressure wave into vibrations of a mechanical element, and detect these vibrations electrically via changes in piezoelectricity \cite{nanofiber}, resistivity \cite{resistivity}, magnetic transduction or capacitance \cite{capacitive}. For many applications, high spatial, temporal and directional resolution is a key requirement. This has driven development towards both ultrasonic frequencies, with their correspondingly short acoustic wavelengths, and microscale sensing devices that are capable of resolving such waves at, near, or beyond their diffraction limit~\cite{Miller2016}. The degradation in acoustic sensitivity that comes hand-in-hand with operation at higher frequencies and with smaller sensing areas presents a major challenge~\cite{Ballantine}.  While, for an acoustic wave propagating through gas, the sensitivity is only fundamentally limited by the random momentum kicks from gas molecules  as they collide with the sensor, all existing acoustic sensors are far from this limit. Their noise floor is, instead, typically dominated by electronic noise. This has motivated recent progress in photonic  acoustic sensors~\cite{Preisser,Guggenheim,Wissmeyer}.

In this article we extend cavity optomechanical sensing to the measurement of acoustic and ultrasonic waves, using a lithographically fabricated device suspended above a silicon chip via thin tethers. By engineering its structure for high acoustic sensitivity, we reach the regime where gas molecule collisions dominate the noise floor. This allows noise equivalent pressures of 8--300 $\mu$Pa$/\sqrt{\rm Hz}$ at a range of frequencies between 1~kHz and 1~MHz.  Compared to acoustic sensors that use similar, but non-suspended, optical cavities and rely on refractive index shifts and static deformations rather than nanomechanical resonances~\cite{Kim2017B}, the peak sensitivity represents a more than  three order-of-magnitude advance. Normalised by device area, it outperforms all previous air-coupled ultrasound sensors by two orders-of-magnitude at ultrasound frequencies from 80~kHz to 1~MHz.

\section*{Results}

\subsection*{Working principles of cavity opto-mechanical acoustic sensing}
    \begin{figure}[!htbp]
 \includegraphics[scale=0.65]{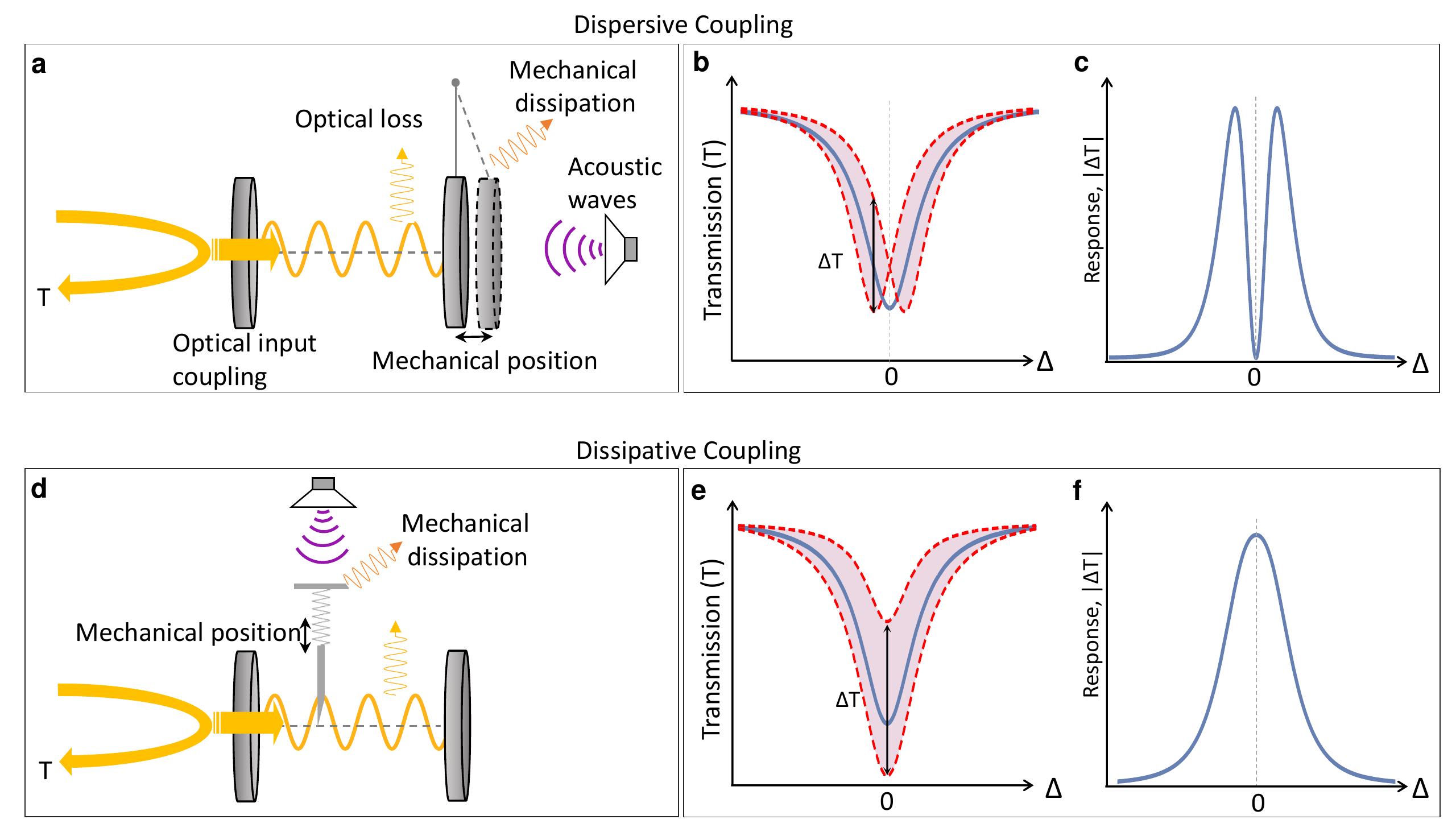}
\caption{{\bf Principles of dispersive and dissipative cavity optomechanical acoustic sensing.} (a\&d) Conceptual schematics of Fabry-P\'{e}rot cavity-based dispersive (a) and dissipative (d) sensors. In (a) an applied acoustic force drives harmonic oscillation of a movable cavity end-mirror modulating the length and resonance frequency of the cavity. In (d) the force drives a mechanical element that modulates the decay rate of the cavity. The modulation is monitored via the change in optical transmission from the cavity. (b\&e) Cavity transmission in the presence of dispersive and dissipative coupling, respectively. The solid blue curves show the cavity transmission for the initial position of the mechanical element while the red dashed curves show the modified cavity transmission due to displacement of the mechanical element. (c\&f) Amplitude of external-force driven modulation in transmission of the cavity optomechanical system for dispersive and dissipative coupling, respectively, versus the detuning $\Delta$ of the input laser field from the cavity resonance. } 
\label{Figure1}
    \end{figure}

In general, cavity optomechanical sensors consist of a mechanically compliant element coupled to an optical cavity. The mechanical element is displaced in response to an external stimulus -- in our case an acoustic wave. The optical cavity resonantly enhances the optical response to this displacement, allowing precise measurement of the stimulus. Commonly, the coupling from displacement to optical response can occur in one of two ways: dispersive~\cite{dispersivecoupling} or dissipative coupling~\cite{Hammerer, Knittel, Li2009}, both of which are used in our sensor. With dispersive coupling, the mechanical displacement alters the cavity length, and therefore optical resonance frequency (See Fig.~\ref{Figure1}a\&b). In dissipative coupling, the displacement instead alters the cavity decay rate, by modifying either the optical input coupling or intra-cavity loss  (See Fig.~\ref{Figure1}d\&e). The concept of ultrasound sensors based on each coupling mechanism is shown in Fig.~\ref{Figure1}, using a Fabry-P\'{e}rot cavity as an illustrative example. In both cases, the output optical signal is linearly proportional to the amplitude of the applied acoustic wave.
 
In the simple case where the mechanical element has a single mechanical resonance, the minimum detectable acoustic pressure for both dispersive and dissipative cavity optomechanical sensing is given by
\begin{equation}
P_{\rm min}(\omega)=\frac{1}{r \zeta A}\sqrt{2 ( \mu l + m \gamma) k_B T+ \frac{1}{N |\chi(\omega,\Delta)|^2}}, \label{P_min}
\end{equation}
 where  $A$ and $T$ are the area and temperature of the sensor, respectively,  and we assume that the laser used to probe the optical response is shot noise limited (see Supplementary Information for derivation). The acoustic pressure wave will only exert a force if it induces a pressure difference  between the top and bottom surfaces of the mechanical element. This is quantified by $r$, the ratio of the pressure difference to the peak pressure at the antinode of the acoustic wave. $\zeta$ is the spatial overlap of the displacement profile of the mechanical sensing element with the incident pressure wave (see Supplementary Information). The first of the three terms under the square-root quantifies the thermomechanical noise introduced by collisions with molecules in the gas surrounding the resonator, where $\mu$ is the coefficient of viscosity of the gas and $l$ is a device geometry-dependent characteristic length-scale.  The second term quantifies the thermomechanical noise introduced by the fluctuation-dissipation theorem due to the intrinsic damping of the mechanical resonator. $m$ is the resonator effective mass which is generally close to but less than the actual mass, and $\gamma$ is the intrinsic mechanical damping rate.
 The third term quantifies the optical measurement noise, with $N$ being the number of photons in the cavity and $\chi(\omega,\Delta)$ an optomechanical susceptibility which accounts for the optical and mechanical response of the sensor as a function of acoustic drive frequency $\omega$ and cavity detuning $\Delta$. In the case relevant to our experiments, where the cavity decay rate is much faster than the acoustic drive frequency, the acoustic frequency dependence is determined solely by the mechanical response and is independent of coupling mechanism. On the other hand, the detuning dependence is fundamentally different for dispersive and dissipative coupling (see Supplementary Information). For optical intensity measurement, $|\chi|$ is zero when the probe laser is tuned to the cavity resonance ($\Delta=0$), and maximised when it is detuned by $|\Delta|=\dfrac{\kappa}{2\sqrt{3}}$. Conversely, for dissipative coupling, $|\chi|$ is generally maximised for on-resonance optical driving. This difference is illustrated in Fig.~\ref{Figure1}b\&e. We finally note that, while derived here for cavity optomechanical sensing, equation~(\ref{P_min}) is applicable quite generally when a mechanical resonance is used to enhance the response of an acoustic sensor in a gaseous environment~(such as~\cite{nanofiber,resistivity,capacitive,Optical-acoustic}) -- only the measurement noise term need be replaced to align with the specific choice of transduction mechanism. 

Fundamentally, the sensitivity of photoacoustic sensing is limited by the thermal energy of the medium through which the acoustic wave propagates. 
In liquids, resonant ultrasound sensors approach to within a factor of two of this thermal limit~\cite{Winkler}.  However, the far lower acoustic impedance of gaseous media greatly reduces both the magnitude of the thermal noise and the efficiency with which acoustic signals can be detected, significantly increasing the challenge~\cite{Kim2017B}.
In this case, the thermal limit results from collisions of gas molecules with the sensor surface, which introduces gas damping of the mechanical energy and
%
%
%
%
%
 is associated with the first term under the square-root in equation~(\ref{P_min}). For the characteristic viscous length-scale of our devices ($l\sim 8$~mm, see Supplementary Information), their area of $A \sim 0.05~$mm$^2$, an ideal pressure participation ratio and spatial overlap ($r=\zeta=1$), and a surrounding gas of air at room temperature ($\mu = 1.8\times10^{-5}\text{ kg/m s}$) we find this gas-damping thermal limit to be $P_{\rm min} \sim 1~\mu\text{Pa Hz}^{-1/2}$. This predicted fundamental-noise limited sensitivity is many orders of magnitude superior to previously reported ultrasound sensors of comparable size~\cite{Guggenheim}. For larger centimetre-scale sensors, the limit drops to tens of nanopascal levels, also well beyond the state-of-the-art. To reach it, the intrinsic mechanical damping rate ($\gamma$) must be smaller than the gas-damping rate ($\gamma_{\rm gas} = \mu l/m$), such that a high quality, low mass, mechanical resonator is advantageous. Furthermore, the measurement noise must be small enough to allow resolution of the random thermal force from collisions of  gas molecules with the resonator. In general, it has proved challenging to simultaneously satisfy  these requirements. However, they align closely with the characteristics of optomechanical devices developed over the past decade to study the quantum physics of nanoscale motion~(see e.g. \cite{Kippenberg spoked toroids,Vahala}).

\subsection*{Sensor design and characterisation}

\begin{figure}[!htbp]
	\includegraphics[width=1 \textwidth]{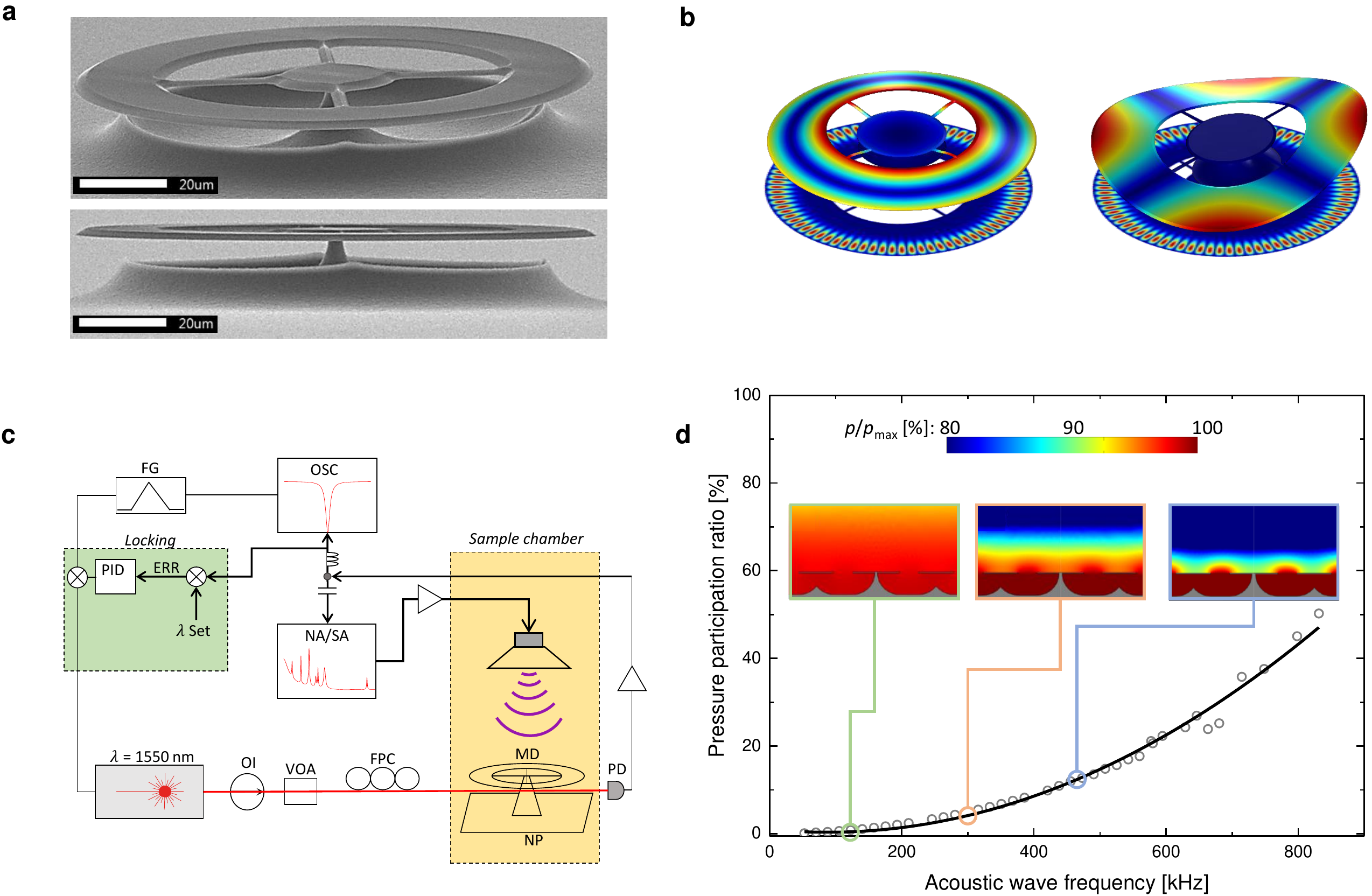}
	\caption{{\bf Device architecture and experimental schematic.} (a) Scanning electron micrograph of the microdisk used in this study. The microdisk is an optical cavity which is evanescently coupled to a tapered optical fibre. (b)
		Finite-element simulations of the modeshapes of two typical mechanical modes  of the microdisk ({\it  left:} second-order flapping mode, {\it right:} crown mode). (c) Shows the phase sensitive and thermally stabilised experimental setup used to characterise the sensor. NP: nanopositioner; MD: microdisk; PD: photodetector; FPC: fibre polarization controller; VOA: variable optical attenuator; OI: optical isolator; FG: function generator; OSC: digital oscilloscope; NA: network analyser; SA: spectrum analyser.  (d) Shows the simulated pressure participation ratio, i.e. the fraction of the total acoustic pressure acting on the mechanical structure, for a number of frequencies. The insets display the pressure distribution at 105, 281 and 421 kHz, respectively; while $p/p_{\rm max}$ is the ratio of the pressure to the pressure at the antinodes of the acoustic wave. } 
	\label{experiment}
\end{figure}

Here we develop a suspended spoked silica microdisk optomechanical system purpose-designed for ultrasensitive ultrasound detection, as shown in Fig.~\ref{experiment}a. Similar to a regular microdisk cavity, light is confined in a high quality whispering-gallery mode around the periphery of the disk, maximising both the optomechanical susceptibility $\chi$ and the intracavity photon number $N$ for a given incident optical power. The use of thin spokes to suspend the disk above a silicon substrate both further increases the optomechanical susceptibility by increasing the compliance of the mechanical structure, and  isolates the mechanical resonances, greatly suppressing the intrinsic mechanical damping~\cite{Kippenberg spoked toroids}. One compromise associated with the use of  spokes is a reduction in active sensing area.    
Here, we optimise the active area within the constraints of the device footprint to functionalise spoked microdisks for efficient ultrasound detection. We find that 
high mechanical compliance and isolation can both be achieved while maintaining a 70\% active area, such that  the reduction in area only minimally influences the acoustic sensitivity. 

While suspension of the mechanical element offers significant advantages in terms of mechanical quality and compliance, a potential disadvantage is that its underside is not isolated from the acoustic pressure wave. One might expect this to reduce the pressure difference across the resonator, decreasing the pressure participation ratio and degrading the acoustic sensitivity. To explore this behaviour, we perform finite-element simulations of an acoustic plane wave incident on a spoked silica microdisk, with results shown in Fig.~\ref{experiment}d. The participation ratio is found to increase roughly quadratically with acoustic wave frequency, exceeding 50\% at frequencies above 800~kHz. We attribute the quadratic dependence firstly to the increasing spatial gradient of the pressure wave with increasing frequency and, secondly, to an increasing resonant confinement of sound between the sensor and the substrate, as the acoustic wavelength becomes comparable to the height of the airgap beneath the sensor.  

The spoked microdisk is photolithographically fabricated with outer and inner radii of 148 $\mu$m and 82 $\mu$m, respectively, and a $\sim \! 1.8~\mu$m device thickness, resulting in a small mass of approximately $230$~ng (see Methods and SEM image in Fig.~\ref{experiment}a). The probe laser is evanescently coupled into, and out of, the microdisk via an optical nanofibre, facilitating direct coupling into fibre-optic systems. We note that on-chip packaging is also possible by replacing the nanofibre with an integrated optical waveguide~\cite{Baker2011}. The microdisk supports families of mechanical eigenmodes that can be resonantly driven via an acoustic field (see Fig.~\ref{experiment}b). The dominant effect of microdisk vibrations on the cavity resonance is generally to modify the resonance frequency, providing a mechanism for dispersive optomechanical sensing. However, vibrations can also enable dissipative sensing, modifying the distance between fibre and microdisk and therefore the cavity input coupling.

Using the experimental setup shown in Fig.~\ref{experiment}c, the mechanical and optical modes of the sensor, as well as its acoustic response, were investigated via their effect on the transmission of the probe laser through the nanofibre. 
An optical cavity mode with wavelength of around 
$\lambda = 1555.7$~nm, in the telecommunications C-band, and with intrinsic quality factor of  $3.6 \times 10^6$ was selected for the experiments (see Supplementary Fig.~1 and Methods). A feedback loop was used to lock the laser wavelength at a fixed detuning with respect to this mode, such that the experiment was insensitive to low-frequency thermal fluctuations in the cavity and optical fibre circuit and drift of the probe laser wavelength.

      \begin{figure}[!htbp]
 \includegraphics[width=0.8 \textwidth]{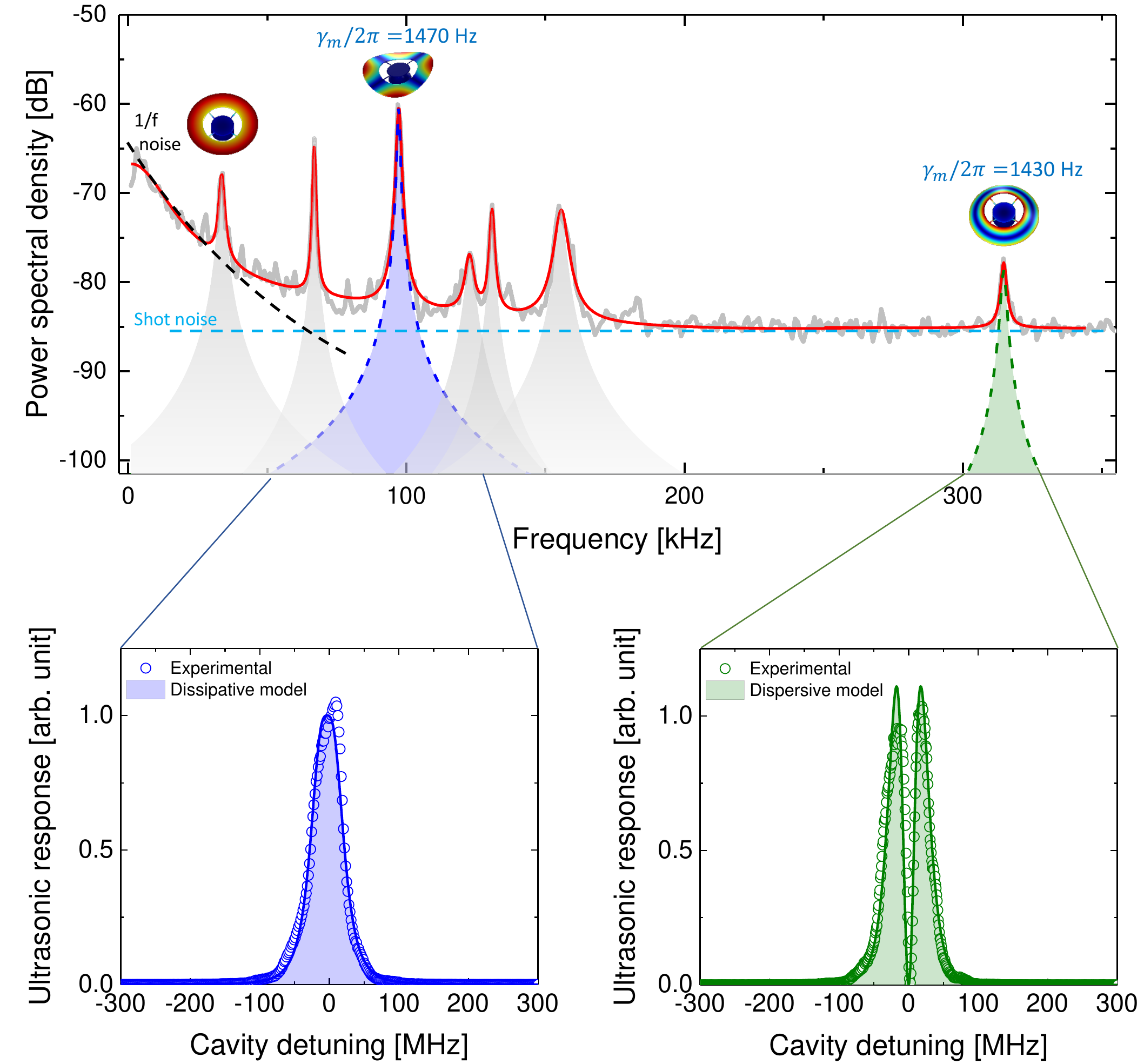}
\caption{{\bf Noise spectrum, mechanical modes, and dissipative and dispersive coupling.} Top panel shows the noise spectral density of the microdisk coupled to the tapered fibre in the absence of acoustic signal. The blue dashed line specifies the shot noise level given by the laser intensity, and the black dashed line corresponds to the 1/$f$ noise. The shaded Lorentzian peaks specify the combined noise due to intrinsic damping and gas damping for several mechanical modes of the device. The green and blue shading highlights examples of dispersively and dissipatively coupled mechanical modes, respectively. $\gamma_m$ quantifies the total mechanical dissipation rate of each of these modes, including both gas and intrinsic damping. The bottom panels show the ultrasonic response as a function of laser-cavity detuning at frequencies of 98 kHz and 315 kHz, resonant with the second-order crown and flapping modes of the disk, respectively. The shaded areas are fits based upon the theoretical expectation for system response as function of detuning (see Supplementary Information) corresponding to dissipative  (left panel) and dispersive (right panel) coupling, respectively. } 
\label{noise}
      \end{figure}

To investigate the response, noise performance and sensitivity of the sensor, we detuned the laser away from the optical resonance to the point of maximum slope with respect to the cavity dispersion, optimising the dispersive transduction of acoustic signals. The noise spectral density of the sensor was then measured using a spectrum analyser, as shown in Fig.~\ref{noise}. At low frequencies ($\lesssim$50~kHz), the dominant noise mechanism is 1/$f$ noise. At higher frequencies, the noise floor is dominated either by laser shot noise or, near the resonance frequencies of mechanical eigenmodes, thermomechanical noise due to the combination of intrinsic and gas damping
  with characteristic sharply peaked Lorentzian frequency response. 

In order to quantitatively verify our model  for cavity optomechanical acoustic sensing (see equation~(\ref{P_min}) and Fig.~\ref{Figure1}c,f), we examined the acoustic response for the second-order crown and flapping modes of the microdisk shown in Fig.~\ref{experiment}b. A piezo-electric element (PZT) was used as an ultrasonic transmitter, creating an ultrasonic wave at each frequency,  and  the response of the sensor was analysed using a vector network analyser. Specifically, the off-diagonal scattering parameter (i.e., the coherent power transmission from the PZT to the photodetector through the sensor) was recorded  as a function of laser-cavity detuning. The results are shown in the bottom panels of Fig.~\ref{noise}. The response of the flapping mode is zero on cavity resonance, with maxima on either side, characteristic of the usual dispersive coupling (c.f. Fig.~\ref{Figure1}f). On the other hand, the crown mode features a maximum at zero detuning, characteristic of dissipative coupling (c.f. Fig.~\ref{Figure1}c). Dissipative coupling has been observed previously in a waveguide-coupled microdisk~\cite{Li2009}. In our case it is most likely due to the large vertical displacement amplitude of the mode which modulates the taper-microcavity separation, combined with first-order suppression of dispersive coupling inherent to crown modes. 
The results show very good agreement  to respective fits to dissipative and dispersive coupling, as shown in Fig.~\ref{noise}, validating the theoretical models for both sensing mechanisms.   
 
\subsection*{Characterising the sensor: Dynamic range and sensitivity}

\begin{figure}[!htbp]
	\includegraphics[width=1 \textwidth]{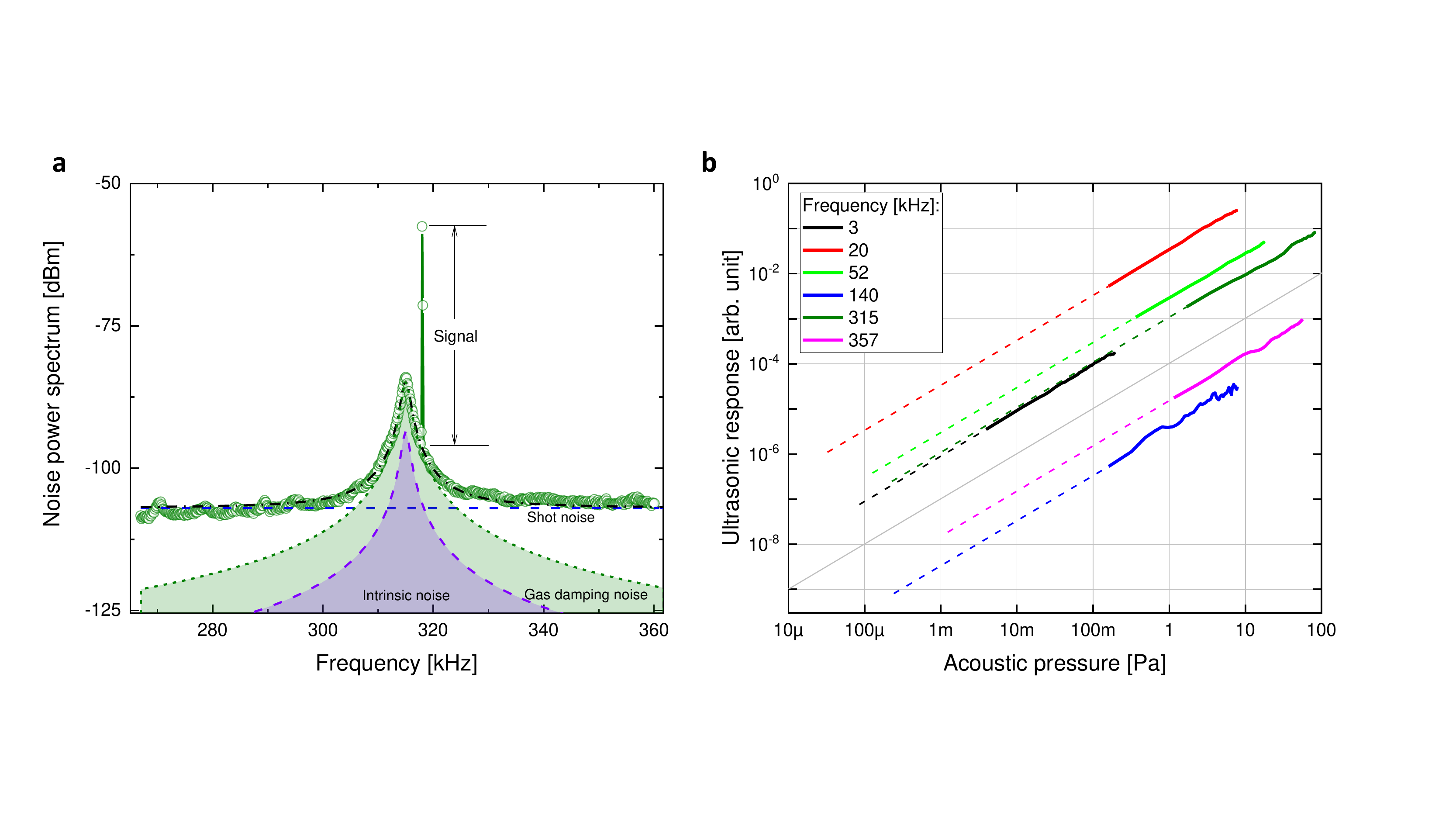}
	\caption{{\bf Evaluation of the noise equivalent pressure sensitivity and the linear dynamic range (LDR).} (a) Noise spectral density of the sensor near a mechanical mode of the microdisk measured at an electrical bandwidth of 200~Hz. An ultrasonic pressure of 120~mPa at frequency of 318 kHz is applied to the device resulting in a signal to noise ratio of $\sim$40~dB. The shot noise is shown with the dashed blue line. 
The thermomechanical noise introduced by intrinsic and gas damping is shown, respectively, by the purple and green shaded Lorenztian's.
The total noise is fitted with the black dash-dot line, in good agreement with theory, and is dominated by gas damping noise between 306~and~325~kHz.
(b) Ultrasonic response of the sensor at different frequencies as a function of ultrasonic pressure. The dashed grey line is a guide to the eye indicating the expected slope for a linear response. The LDR is $>$120 dB for a measurement integration time of 1 second, with its upper limit dictated by  the measurement setup rather than the acoustic response. The solid lines correspond to the measured data for each frequency and the dashed lines connect these to the noise equivalent pressure that sets the lower limit of the LDR. }
	\label{LDR}
\end{figure}

To experimentally quantify the noise equivalent pressure of the sensor, we interferometrically calibrated the displacement of the PZT element as function of its drive frequency. The acoustic pressure generated by the PZT was calculated from its displacement, air acoustic impedance and its distance to the sensor (see Supplementary Information for details). The ultrasonic response of the system was then measured at different frequencies for which the applied pressure was known. Fig.~\ref{LDR}a shows, as an example, the response at 318~kHz in the wing of the second-order flapping mode, relative to both the shot noise and thermomechanical noise introduced by intrinsic and gas damping.
 The signal-to-noise ratio is $\rm{SNR} \sim 40$~dB with an applied pressure of $P_{\text{applied}} = 120$~mPa, measured over an integration time of $\tau = \Delta f^{-1}$ where $\Delta f = 200$~Hz is the spectrum analyser resolution bandwidth. The noise equivalent pressure can then be calculated as
\begin{equation}
P_{\rm min}(\omega)=\sqrt{\frac{\tau}{\rm{SNR}}} \times P_{\text{applied}}(\omega) \sim 84~\mu \text{Pa}/\sqrt{\text{Hz}}.
\end{equation}
This is in reasonably good agreement with the thermomechanical noise-dominated noise equivalent pressure predicted from equation~(\ref{P_min}) of
$100~\mu \text{Pa}/\sqrt{\text{Hz}}$, 
given the device area and temperature ($T=300$~K), our simulated pressure participation ratio at 318~kHz of $r=0.055$, the effective mass of the flapping mode of $m=110$~ng, its measured total mechanical damping rate $\gamma_{\rm m}/2\pi= (\gamma + \gamma_{\rm gas})/2\pi=1,430$~Hz, and its spatial overlap $\zeta=0.14$ with a plane pressure wave (see Supplementary Information). 

It is informative to examine the contributions to the 
noise equivalent pressure from intrinsic mechanical dissipation, optical shot noise and fundamental gas damping. As can be seen from Fig.~\ref{LDR}a, at the second-order flapping mode resonance frequency the shot noise power spectrum is 13 dB below the combined thermomechanical noise from gas and intrinsic damping, contributing 5\% in power to the total noise. The fluctuation dissipation theorem dictates that the ratio of noise power introduced by gas damping and intrinsic mechanical damping is equal  to the ratio of the damping rates, as may be directly confirmed from equation~(\ref{P_min}). 
By measuring the mechanical damping rate of the  flapping mode as a function of background pressure, we  isolated these two components, finding that $\gamma_{\rm gas}/2\pi=1,260$~Hz and $\gamma/2\pi=170$~Hz (see Supplementary Information). 
The gas damping noise power therefore dominates by a factor of $\gamma_{\rm gas}/\gamma \sim 7.4$. All-in-all, these results show that, at this acoustic frequency, the noise equivalent pressure of the sensor is within 9\% of the noise floor introduced by thermal collisions of gas molecules with the sensing element. This gas damping noise floor is fundamental, in that it cannot be eliminated without removing the gas through which the acoustic wave itself propagates. To our knowledge, our sensor is the first acoustic sensor which is sufficiently sensitive for it to dominate.
 
The resonantly enhanced bandwidth of the sensor around the second-order flapping mode is given by the frequency band where the combined thermomechanical noise dominates shot noise, i.e. between 306~and~325~kHz.
The noise equivalent pressure is relatively constant over this frequency range,  before degrading at frequencies further from resonance. This bandwidth could, in future, be extended by increasing the optical power used to probe the sensor (and therefore $N$ in equation~(\ref{P_min})) or even by using quantum correlations to reduce the optical noise level for fixed optical power~\cite{Beibei}.
 
To explore the wider bandwidth, we measured the response and noise equivalent pressure for acoustic waves over the frequency range from 1~kHz to 1~MHz (see Supplementary Information Figs.~7~\&~8).  As expected for a resonant sensor, both parameters vary significantly over this range, exhibiting sharp resonant features.
Resonantly enhanced narrowband sensitivities of 8--300~$\mu$Pa/$\sqrt{\rm{Hz}}$ are achieved for many frequencies across the range, with a broadband sensitivity better than 10~mPa$/\sqrt{\rm{Hz}}$ maintained at all measured frequencies.
The upper limit of 1~MHz is not intrinsic, but rather introduced by the inability to generate acoustic waves at higher frequencies due to the frequency response of our PZT transducer and the high acoustic attenuation of air at high frequencies. Indeed, mechanical resonances at hundred megahertz frequencies have been observed in cavity optomechanical systems of similar size to those reported here~(see e.g.~\cite{Schliesser2008}); while gigahertz resonance frequencies are available in smaller devices~(see e.g.~\cite{Ding2010}). Consequently, our approach can be expected to perform well into this higher frequency range. 
While the device was not optimised for audio frequencies, its performance at these lower frequencies  remained sufficient to record the Chris Jones song ``Long After You're Gone'' in the lab environment by digitizing the output of the photo-detector with no further processing and filtering (see online Supporting Information, Video 1). 

To investigate how the ultrasonic response changes when varying the magnitude of the acoustic pressure, we recorded the system response at various frequencies as a function of the applied pressure.
As shown in Figure \ref{LDR}b, the sensor has a linear dynamic range (LDR) of 120~dB.
The lower bound on the linear dynamic range (LDR) of any sensor is given by the noise equivalent signal (pressure sensitivity in case of an acoustic sensor) and the upper bound is the deviation point from linearity~\cite{LDR}. In our experiments, this upper limit is set by the maximum accessible pressure of $\sim$100~Pa, with the sensor response linear throughout the range at all tested frequencies. Hence, the reported LDR is an underestimation.

\subsection*{Discussion}

\begin{figure}[!htbp]
 \includegraphics[width=0.6 \textwidth]{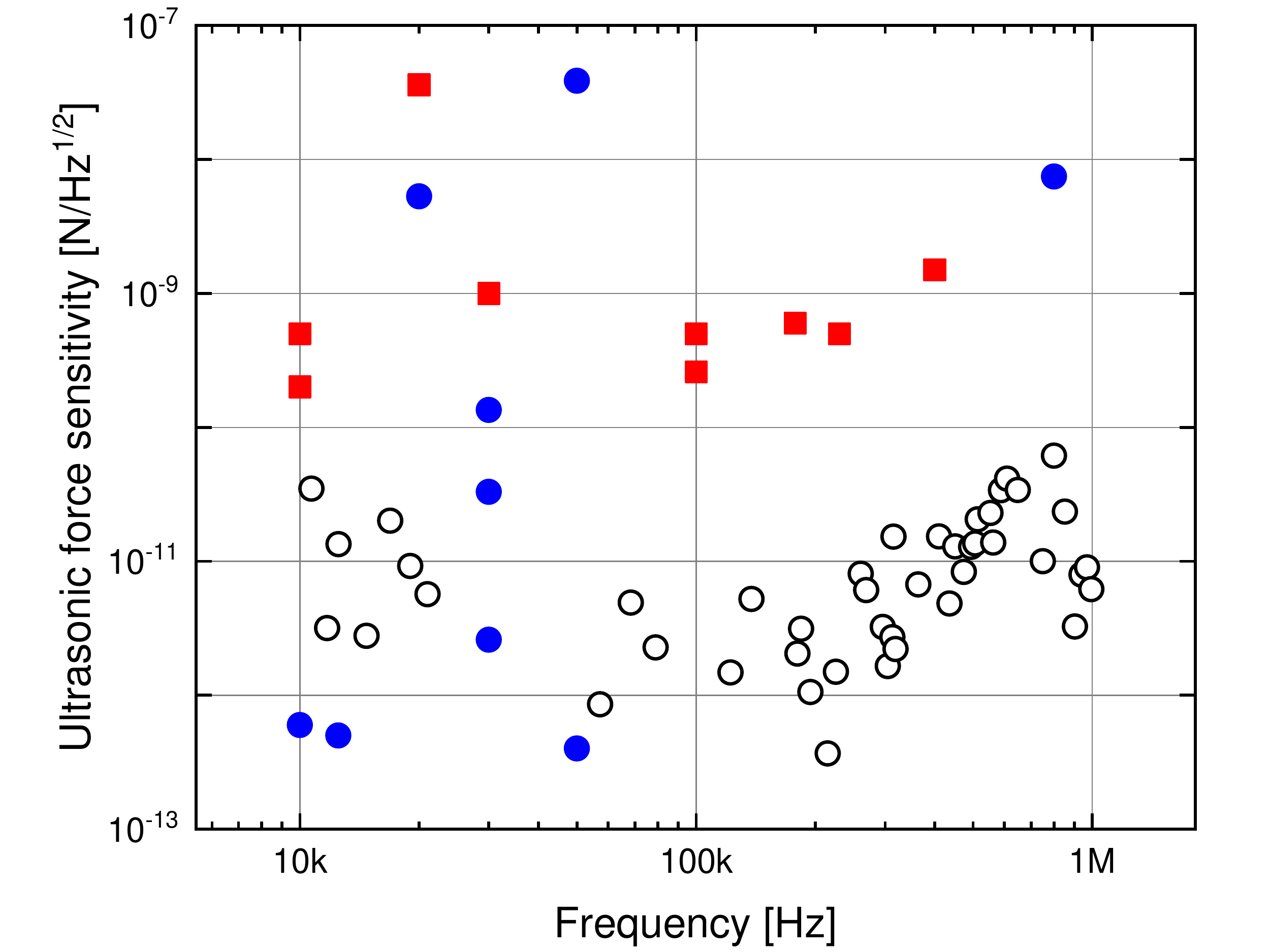}
\caption{{\bf Ultrasonic force sensitivity in comparison with other air-coupled sensors}. Ultrasonic force sensitivity is evaluated  as the noise equivalent pressure sensitivity multiplied by the sensing area and plotted versus frequency: open circles correspond to this work and  solid symbols show results of other optical (blue circles) and electrical (red squares) approaches. The improvement of the sensitivity in this work is especially notable between  80~kHz to 1~MHz. Citations to previous work are provided in the Supplementary Information. }
\label{stats}
\end{figure}

It is interesting to compare the sensitivity of our sensor with existing ultrasound sensors. The peak sensitivity represents a more than three order-of-magnitude advance on previous comparable air-coupled optical sensors~\cite{Kim2017}, and is competitive with the best liquid-coupled piezoelectric sensors~\cite{Winkler} which benefit from four orders-of-magnitude larger sensing area and near-ideal acoustic impedance matching.

The force experienced by an ultrasound sensor
scales linearly with sensing area. Consequently, as a general rule, sensitivity  improves as the sensing area increases. To compare our sensor to ultrasound sensors of different sizes, we therefore calculate the {\it ultrasonic force sensitivity}, normalising the pressure sensitivity to area. Fig.~\ref{stats} shows the comparison  to other air-coupled sensors over the frequency range from 10~kHz to 1~MHz.
 The performance is particularly  good  at frequencies between 80~kHz and 1~MHz, where the ultrasonic force sensitivity represents an advance of  approximately two orders-of-magnitude. While this demonstrates that the sensor is an especially good acoustic {\it force} sensor, it
is worth noting that while the absolute {\it pressure} sensitivity in equation~(\ref{P_min}) does include an explicit inverse-area scaling, it also includes implicit dependence on area through other parameters, such as the sensor mass and characteristic length-scale $l$. Consequently, the comparison between sensors of different size cannot be straightforwardly extended to absolute sensitivity.
 
Compared to liquid-coupled sensors, the peak ultrasonic force sensitivity of $370\,$fN$\,$Hz$^{-1/2}$ is more than three-orders-of-magnitude superior to state-of-the-art piezoelectric sensors~\cite{Wissmeyer}, while also offering somewhat improved broadband sensitivity. The peak force sensitivity also exceeds optical liquid-coupled sensors, such as the Fabry-P\'{e}rot sensor in Ref.~\cite{Preisser} which has sensitivity of around $1.8$~nN$\,$Hz$^{-1/2}$ (see Supplementary Information) and microring sensors operating at $1.8$~pN$\,$Hz$^{-1/2}$~\cite{Wissmeyer}. That the sensor is comparable, both in terms of absolute pressure sensitivity and force sensitivity, with liquid-coupled sensors is notable given the large reduction in acoustic energy transport at the air-sensor interface due to the more than three orders-of-magnitude lower acoustic impedance of air compared with liquids~\cite{Kim2017}.
 %
 %
 %

%
%
%
%

The sensor could be scaled straightforwardly to larger or smaller sizes, for improved absolute pressure sensitivity or improved resolution/high frequency sensitivity, respectively. The sensitivity could be further improved by engineering the physical structure of the device to increase the pressure participation ratio, decrease the noise contribution from collisions with thermal gas molecules, and improve the overlap of the mechanical motion with the incident pressure wave. The participation ratio can be optimised, for a given acoustic wave frequency, by controlling the height of the sensor above the silicon substrate. Indeed, our modelling suggests that participation ratios even exceeding $r=1$ are achievable due to resonant enhancement of the pressure wave between the substrate and device.  This, in effect, would represent a microscale acoustic resonator fabricated on a silicon chip, with significant advantages over the bulk-machined acoustic resonators often used to enhance acoustic pressure waves in other approaches~\cite{Dong}. The overlap $\zeta$ could be increased to near unity by engineering the resonance frequency of a suitable mechanical mode, such as the first order flapping mode, to coincide with the frequency of the pressure wave. The noise contribution from thermal gas molecule collisions is determined by the geometry-dependent characteristic length-scale $l$, which includes the effects of both squeeze-film molecular damping  and air-drag damping. 
Squeeze-film damping arises from the gas trapped between the device and the substrate, and scales as inverse-height cubed. We estimate that it dominates air-drag damping by a factor of twenty for our current device design (see Supplementary Information), degrading the gas-damping-limited sensitivity by around a factor of five. 
By increasing the separation of the device from the substrate to both suppress squeeze-film damping and enhance the participation ratio, 
the sensitivity could be improved by more than a factor of one hundred, reaching the sub-micropascal regime.


The improved ultrasound sensitivity and microscale resolution offered by our new acoustic sensing technique has prospects for a range of applications. For instance, it could allow 
 improved navigation and spatial imaging in unmanned and autonomous vehicles~\cite{Schmid}; and higher sensitivity high resolution photoacoustic trace gas sensing~\cite{Dong}. In trace gas sensing, the sensitivity reported here could allow detection of carbon dioxide at ten-part-per-billion concentrations with unprecedented spatial resolution (See Supplementary Information). This could, for example, enable measurements of the respiration of individual cells and bacteria, such as photosynthesis and gas exchange through the cell membrane~\cite{PASleaf,Oswald}. Our sensor could also be applied to observe acoustic waves generated by the nanoscale vibrations associated with cellular metabolism~\cite{Longo}. Measurements of these vibrations have been shown to allow diagnostic assays of cellular toxicity and antibiotic resistance~\cite{Longo}, and provide insight into molecular processes such as conformational changes~\cite{Alonso-Sarduy}. Unlike current atomic force microscope-based approaches \cite{Longo}, our sensor could allow these measurements to be performed without physical contact, and therefore without disrupting the observed processes or contaminating the sensor. Moreover, the measurements could be performed with higher bandwidth, and resolve 100-picometre-level cellular vibration amplitudes at low kilohertz frequencies and sub-picometer vibrations at above 100 kHz (see Supplementary Information).

As with all ultrasonic sensors that use mechanical resonances~(e.g.~\cite{nanofiber,resistivity,capacitive}), one potential drawback of our sensor is that the best sensitivity is only achieved in narrow frequency windows near each mechanical resonance. This is not a concern for applications such as trace gas sensing and narrowband sonar where the signal is an acoustic tone of known frequency. In scenarios where broadband sensitivity is required, our approach has several attractive features compared to other resonant sensors. Firstly, the sensor is able to operate simultaneously on multiple mechanical resonances over the full 1~kHz to 1~MHz frequency band. Secondly, the combination of optical measurement and cavity enhancement provides a low shot noise floor, allowing high sensitivity even away from resonance. Finally, the cavity optomechanical architecture allows the use of techniques from quantum optomechanics 
to enhance the broadband response of future sensors~\cite{BowenMilburn}. For instance, the optical shot noise could be suppressed by engineering the cavity structure to increase the optomechanical coupling (such as in, e.g.,~\cite{Vahala}) or using quantum correlated light~\cite{Beibei},  laser cooling could be used to broaden and flatten the mechanical resonances without introducing additional thermal noise~\cite{Chan,Teufel,Schliesser,Bowen,Harris,Kim2017} (see Supplementary Information), or laser levitated particles could be used to entirely remove substrate thermal noise~\cite{Geraci}.

\subsection*{Methods}    
  
{\bf Device fabrication}: The spoked microdisks were fabricated on silicon wafers, covered with a 1.8~$\mu$m layer of thermally grown silicon dioxide. The wafer was first coated with photoresist and spoked circular pads were defined using UV-photolithography (see Supplementary Fig.~6a). After developing the photoresist, the wafer was exposed to buffered Hydrofluoric acid, removing all the uncovered silicon dioxide (see Supplementary Fig.~6b)). The remaining photoresist was consecutively cleaned off with acetone (see Supplementary Information Fig.~6c). In the subsequent step, the wafer was coated again with photoresist for protection and mechanically separated into about thirty chips containing ten circular silicon dioxide structures each. After separation, the photoresist was removed and the chips were individually exposed to XeF$_2$ gas, selectively removing the silicon and releasing the silica structures (see Supplementary Fig.~6d and e).

{\bf Characterisation setup}: Light from a 1555 nm tunable Erbium-doped fibre laser [NKT Photonics, Koheras Adjustik] was guided to the experiment through an optical isolator to avoid reflection back into the laser. The intensity of the laser was adjusted using a variable fibre attenuator. The frequency of the laser could be thermally tuned over a range of about one nm or electronically swept over tens of picometers using a built-in piezo element of the laser cavity. The polarization of the light was adjusted using a fibre polarization controller (FPC). A tapered nanofibre was used to evanescently couple the laser to a whispering-gallery-mode of the disk. The crucial coupling distance between the taper and the disk was coarsely adjusted using manual micrometer stages and optical microscopes. Fine-tuning was implemented using a nanopositioning stage [Thorlabs MDT693A]. The transmitted light had intensity of around 20~$\mu$W and was detected with an InGaAs-photodetector [New Focus 1811 DC-125MHz]. 

{\bf Optical mode characterisation}: The optical mode of the microdisk was investigated by measuring the transmission of the probe laser as a function of the laser frequency. The frequency of the laser ($\lambda = 1555.716$ nm) was swept over the cavity optical mode using a function generator (FG) and the probed laser transmssion was recorded with an oscilloscope (OSC) (see Fig.~\ref{experiment}c). The optical mode was found to have a quality factor $Q =1.8 \times 10^6$ when the tapered fibre was positioned so that the input optical coupling rate matched the intracavity loss rate, i.e. critical coupling (see Supplementary Fig.~1). This implies an intrinsic quality factor of $3.6 \times 10^6$.

{\bf Noise floor and signal response}: The high-frequency part of the signal was Fourier transformed in a spectrum analyser [Agilent N9010A] to
analyse the sensor noise spectrum, and to calibrate the sensor signal to noise ratio (SNR). The system network response was measured using a vector network analyser [Agilent E5061B] to determine the dependence of the SNR on applied acoustic pressure and to determine sensitivity as a function of frequency. The Network analyser was also used to calibrate a piezo element (Thorlabs AE0505D08F) as a function of frequency and voltage to operate as the acoustic source (see Supplementary Information for detail).

\subsection*{Author contributions}
WPB provided overall leadership for the project. SB, AA and WPB conceptualized the idea and designed the experiments.  SB and AA set up the experiments, performed the measurements and analyzed the data. All co-authors interpreted the data. SF and SB fabricated the devices. SB and WPB developed the theoretical model. SF performed the finite element simulations. All co-authors contributed in the development of the manuscript which was initially drafted by SB and AA.
\subsection*{Acknowledgments}
Authors would like to thank Glen Harris and Xin He for their help with experiments and Chris Baker for his help with finite element simulations. The project was funded primarily by the Australian Research Council through the Discovery Project DP140100734, and a grant through the Air Force Office of Scientific Research. WPB acknowledges an Australian Research Council Future Fellowship (FT140100650). SB is a S\^{e}r Cymru II COFUND Fellow and acknowledges funding from the European Unionâ Horizon 2020 research and innovation programme under the Marie Sklodowska-Curie grant agreement No 663830. AA is a S\^{e}r Cymru II Rising Star Fellow. 

\subsection*{Data availability}
The data that support the findings of this study are available within the paper and its supplementary information files.

\subsection*{Additional information}

\section*{\fontsize{14}{10}\selectfont Supplementary information}

\vspace{2cm}

\section{Characterising the optical resonance} 

\setcounter{figure}{0}
 \begin{figure}[!htbp]
 \includegraphics[scale=0.9]{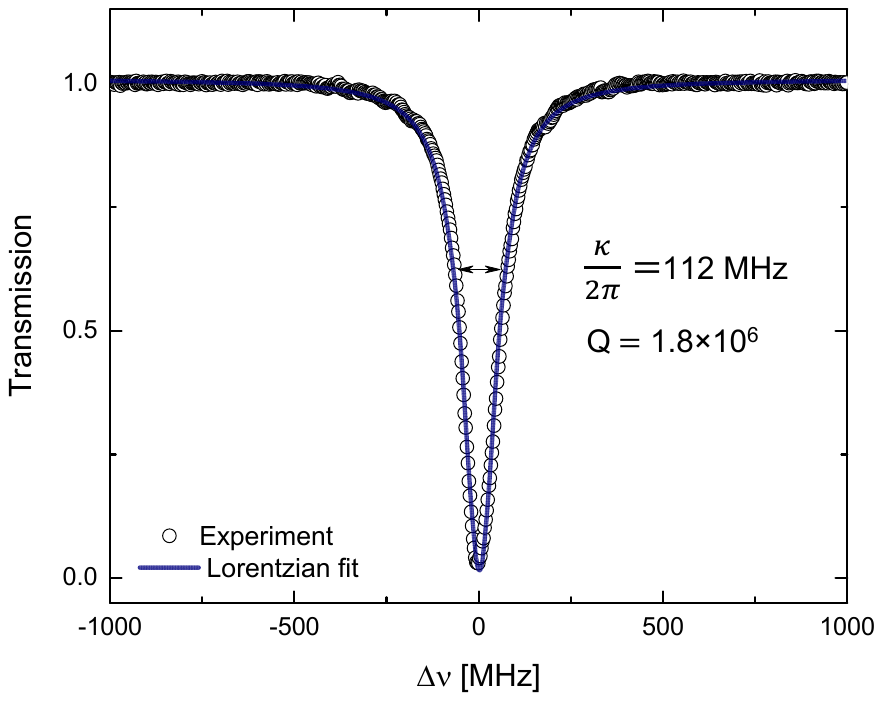}
\caption{{\bf Optical resonance.} One of the optical modes of the micro-disk ($\lambda = 1555.716$ nm) is shown with optical quality factor and cavity damping indicated on the figure. } 
\label{optical_mode}
    \end{figure}
  
The optical resonance used in the experiments was characterised by scanning the frequency of the laser across the mode and fitting the observed transmission through the tapered fibre to an inverted Lorentzian (see Fig.\ref{optical_mode}). This allowed the coupled cavity decay rate $\kappa$ and quality factor $Q$ to be determined. These were found to be $\kappa=112$~MHz and $Q=1.8\times10^6$ in critical coupling regime corresponding to an intrinsic quality factor of $3.6\times10^6$.

\section{Previous ultrasound sensors}

\subsection{References for Fig. 5 in the main text}
\begin{figure}[!htbp]
 \includegraphics[width=0.6 \textwidth]{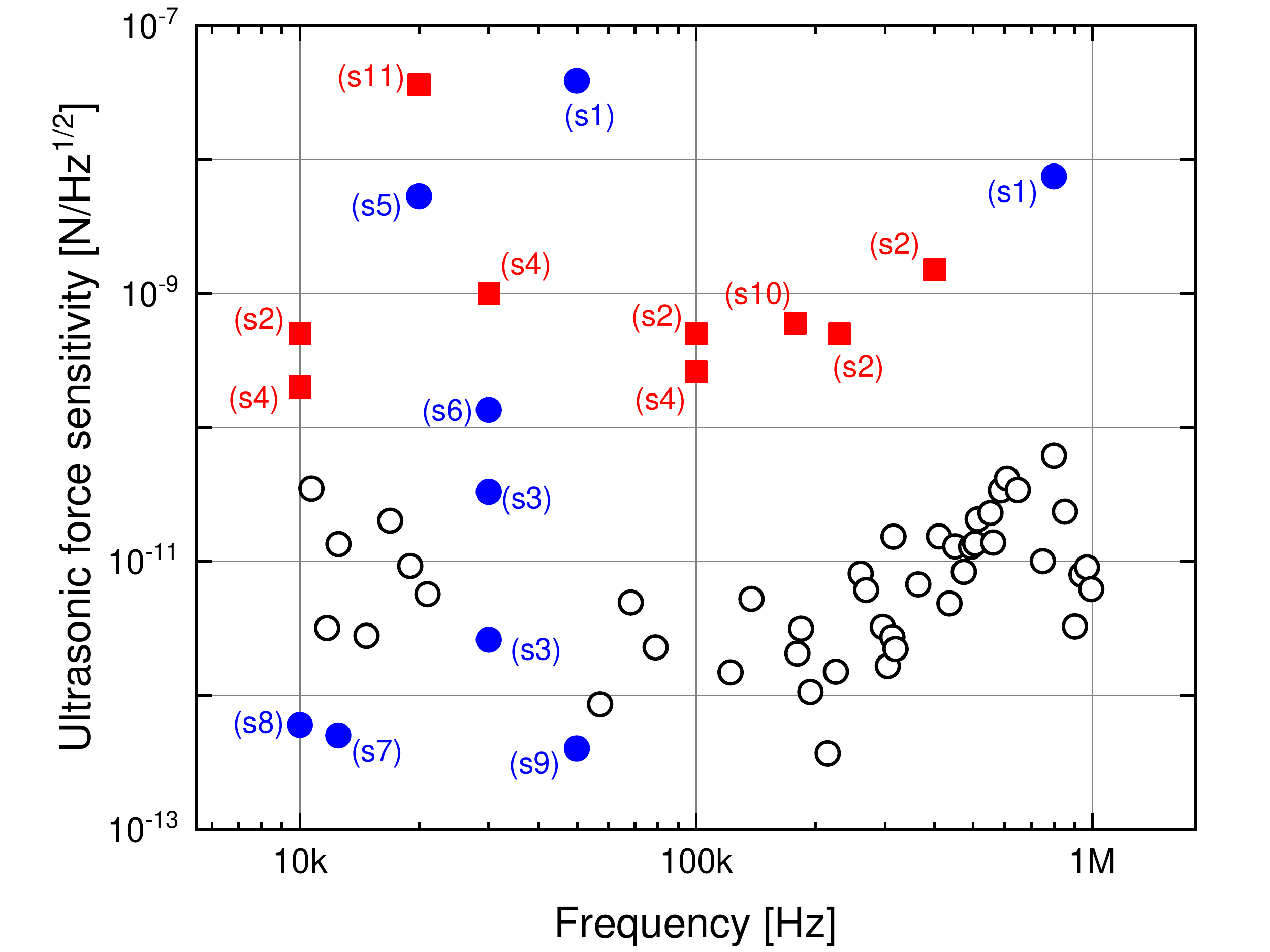}
\caption{{\bf Ultrasonic force sensitivity in comparison with other air-coupled sensors.} s1 to s11 refer to Supplementary Information references \cite{Kim}--\cite{Bucaro}.} 
\label{Supporting_fig5}
    \end{figure}

 Fig.~\ref{Supporting_fig5} provides citations to previous works on acoustic sensors.

\newpage
 \subsection{Effective area and ultrasonic force sensitivity of Fabry-Perot style acoustic sensor}

Apart from the Fabry-Perot style acoustic sensor of~\cite{Preisser}, the area used to normalise the ultrasonic force sensitivity of the sensors discussed in the main text can generally be unambiguously defined. In the case of~\cite{Preisser}, however, the different sensing mechanism, optical detection of the refractive index modulation the pressure wave induces in a gas, makes the appropriate definition less clear. To clarify this, here we consider the diffraction of an acoustic wave incident on their sensing head.

The sensor head in~\cite{Preisser} consists of a semi-enclosured space, with two vertical surfaces serving as mirrors to define the Fabry Perot cavity and two horizontal surfaces consisting of spacers to support the cavity. The final two faces of the enclosure are left open. The system is then immersed in liquid, with an applied plane acoustic wave propagating into the sensing region (the locality of the optical field within the cavity) through the open faces. 
 The acoustic wave modifies the refractive index of the enclosed gas and therefore the optical path length in the cavity. The total field of view for the acoustic wave is approximately 2~mm by 2~mm. The wave propagates for roughly 2~mm within the enclosure before reaching the axis of the laser beam.  

We would like to know how the amplitude of the acoustic wave reaching the axis of the laser beam, and therefore the signal-to-noise, would change due to diffraction if, rather than a plane acoustic wave covering the full field of view of the sensor, the acoustic wave was concentrated on a smaller area of the outer surface of the spacer. If that area matches the cross sectional area of the laser beam through which the acoustic wave propagates (2 mm by 60 micron) and it is found that the amplitude is approximately unchanged by diffraction, the laser beam cross section would be the appropriate sensing area to choose. On the other hand, if diffraction significantly decreases the amplitude of the acoustic wave when it is concentrated to an area matching the laser beam cross section, the appropriate sensing area can be found by increasing the concentration area until there is no significant diffraction. Put another way, were the plane incident wave reduced in area, then for areas for which diffraction is small, this would leave the pressure at the sensing region, and therefore sensitivity, roughly unchanged. However, once the the incident plane wave area is reduced to the point where diffraction plays a significant role, the pressure at the sensing region would decrease for a fixed incident intensity, degrading the sensitivity. 

To estimate the diffraction within the spacer we consider diffraction of a wave with Gaussian profile, noting that this gives a minimum possible diffraction (e.g. the perhaps usual square-profile would diffract faster). The diffraction length is then quantified by the Rayleigh length
\begin{equation}
z_R = \frac{\pi w^2}{\lambda},
\end{equation}
where $w$ is the radius of the acoustic wave incident on the outside of the spacer and $\lambda$ its wavelength. Assuming that the liquid in which the sensor is immersed is water, 
the longitudinal sound velocity is $v=1,500$~m/s. 
For their 1~MHz acoustic wave frequency, this gives a wavelength of $\lambda = v/f = 1.5$~mm. We then ask, what radius of acoustic wave would be required for it to not diffract significantly over the 2~mm propagation distance to the laser beam axis? 
This is given by setting the Rayleigh length equal to 2~mm, so that
\begin{equation}
w = \sqrt{\frac{z_R \lambda}{\pi}} \sim 1~{\rm mm},
\end{equation}
or an acoustic wave diameter of 2~mm. This implies that even an acoustic wave fully spanning the  2 mm by 2 mm field of view of the sensor would experience significant diffraction propagating through the spacer, and indeed given that the incident pressure wave profile will not be Gaussian, that the effective area of the sensor is likely to be larger than the field of view.
To be conservative, in Fig. 5 of the main text we choose the effective area to match the field of view.

Given the reported noise equivalent pressure of 0.45 mPa/Hz$^{1/2}$ in Ref.~\cite{Preisser}, we then arrive at an ultrasonic force sensitivity of 1.8~nN/Hz$^{1/2}$.

    
\section{Derivation of the noise equivalent pressure sensitivity} 
 We start by modelling the motion of a single mode mechanical oscillator at room temperature in response to an external acoustic drive, and probed by a coherent field. In regime where the mechanical thermal noise dominates on resonance, and the quantum back-action noise on the sensor is negligible, we can take the high temperature limit where $\bar{n}(\omega)=k_{\rm B}T/\hbar \omega$. In this regime, we obtain an optical shot-noise limited noise force floor. The Langevin equations of motion for the mechanical displacement and the optical cavity mode respectively are written as
\begin{equation}
m\frac{d^2x_{\rm m}(t)}{dt^2}+m\gamma\frac{dx_{\rm m}(t)}{dt}+kx_{\rm m}(t)=F_{\rm T}+F_{\rm D}(t),
\label{mech-motion}
\end{equation}
\begin{equation}
\frac{da(t)}{dt}=-\frac{i}{\hbar}[a,H_{\rm{sys}}]-\frac{\kappa}{2}a(t)+\sqrt{\kappa_{\rm{in}}}a_{\rm{in}}+\sqrt{\kappa_{\rm l}}a_{\rm l},
\label{Langvin_opt}
\end{equation}
in which $k$ is the spring constant, $F_{\rm T}=\sqrt{2m\gamma k_{\rm B}T}$ is the thermal force, $\gamma$ is the mechanical damping rate, $F_{\rm D}(t)=r\zeta P_{\rm D}(t)A$ is the acoustic drive force in which $r$ is the pressure participation ratio defined in the main text, $P_D$ is the acoustic pressure, $A$ is the sensing area, $\zeta$ quantifies the overlap of the displacement profile of the mechanical sensing element with the incident pressure wave, and  $m$ is the effective mass of the mechanical mode. $\kappa_{\rm in}$ is the input coupling of the cavity to the input optical field, $\kappa_{\rm l}$ is the intrinsic cavity loss and $\kappa=\kappa_{\rm in}+\kappa_{\rm l}$. Moreover, $H_{\rm sys}=\hbar \Delta_{\rm c} a^\dagger a+\hbar g_0a^\dagger ax_{\rm m}$ is the Hamiltonian of the system in the interaction picture rotating with the frequency of the laser in which the optical detuning $\Delta_{\rm c}=\Delta+g_{\rm disp}x_{\rm m}+O^2+...$, includes the dispersive coupling due to the presence of the mechanical oscillation, and $\Delta$ is the optical detuning in absence of the mechanical oscillations. $\kappa_{\rm in}=\kappa_{{\rm in},0}(1-g_{\rm diss} x_{\rm m})$ includes the dissipative coupling of the input field to the cavity in response to the acoustic field up to the first order in $x_{\rm m}$.  $a_{\rm in}$ and $a_{\rm l}$, respectively, show the input optical field into the cavity and the vacuum input noise. In case where the input optical field is a semi-classical coherent laser field, we can displace the amplitude of the optical field such that $a\rightarrow \bar{a}+\alpha_{\rm in}$ where $\vert \alpha_{\rm in}\vert^2=N$ is the input photon intensity. 

The solution to equation (\ref{mech-motion}) in Fourier transformed frequency domain is
\begin{equation}
x_{\rm m}(\omega)=\chi_{\rm m}(\omega)[F_{\rm T}+F_{\rm D}(\omega)],
\end{equation}
in which the mechanical susceptibility $\chi_{\rm m}$ is calculated as $\chi^{-1}_{\rm m}=m(\omega^2_{\rm m}-\omega^2-i\gamma_{\rm m}\omega)$, where $m$ and $\omega_{\rm m}$ are respectively the mass and the resonance frequencyof the mechanical object. The output cavity mode, $a_{\rm out}$ is related to the cavity mode, $a$, and the input mode into the cavity, $ a_{\rm in}$, through the input-output relation \cite{QNoise,WallsMilburn}, $a_{\rm out}=\sqrt{\kappa_{\rm in}}a-a_{\rm in}$. By solving equation (\ref{mech-motion}), the motional displacement of the mechanical resonator can be calculated. Moreover, equation (\ref{Langvin_opt}) can be solved in the frequency domain in linearised displacement regime to get the cavity mode $a$. Using the solutions to equations (\ref{mech-motion},{\ref{Langvin_opt}) together with the input-output relation, the output field of the cavity is calculated as
\begin{equation}
a_{\rm out}(\omega)=\left ( B(\omega) -C(\omega) \right ) x_{\rm m}(\omega)+D(\omega)a_{\rm in}+E(\omega)a_{\rm l},
\end{equation}
in which 
\begin{eqnarray}
&&B(\omega)=\frac{-2i\alpha_{{\rm in}}g_{\rm disp}\kappa_{\rm{in},0}}{(\kappa_0+2i\Delta)(\kappa_0+2i(\Delta-\omega))},\\\nonumber
&&C(\omega)=\frac{2\alpha_{{\rm in}}g_{\rm diss}\kappa_{\rm{in},0}}{\kappa_0+2i(\Delta-\omega)}\Big(1-\frac{2\kappa_{\rm{in},0}}{\kappa_0+2i\Delta}\Big),\\\nonumber
&&D(\omega)=\frac{\kappa_{{\rm in},0}-\kappa_l-2i(\Delta-\omega)}{\kappa_0+2i(\Delta-\omega)},\\\nonumber
&&E(\omega)=\frac{\sqrt{\kappa_{{\rm in},0}\kappa_l}}{\kappa_0+2i(\Delta-\omega)},
\end{eqnarray}
in which $\kappa_{\rm in,0}$ is the original value of the input coupling in absence of the acoustic pressure, $\kappa_0=\kappa_{\rm in,0}+\kappa_{\rm l}$, $g_{\rm disp}=\dfrac{d\Delta}{dx}$ is the dispersive coupling rate and $g_{\rm diss}=\dfrac{1}{\kappa_{\rm{in},0}}\dfrac{d\kappa_{\rm{in}}}{dx}$ is the dissipative coupling rate. Hence, in the regime where $\vert\alpha_{\rm out}\vert\gg\vert\bar{a}\vert$, the output intensity of the cavity can be calculated as $I_{{\rm out}}(\omega)\sim \alpha^*_{\rm out}a_{\rm out}(\omega)+\alpha_{\rm out}a^\dagger_{\rm out}(-\omega)$ where $\alpha_{\rm out}=\vert \alpha_{\rm out}\vert e^{i\varphi}$ is the average amplitude of the output field. The intensity can be rewritten as $I_{{\rm out}}(\omega)\sim \vert \alpha_{\rm out}\vert X^\varphi_{\rm out}(\omega)$, in which $X^\varphi_{\rm out}(\omega)$ is defined as the amplitude quadrature of the output field fluctuations as $X^\varphi_{\rm out}(\omega)=a_{\rm out}(\omega)e^{-i\varphi}+a^\dagger_{\rm out}(-\omega)e^{i\varphi}$. As for the rest of the calculations we require the output fluctuations, we normalize the output intensity as $I_{\rm out}(\omega)\rightarrow\dfrac{I_{\rm out}(\omega)}{\vert\alpha_{\rm out}\vert}\sim X^\varphi_{\rm out}(\omega)$.
For the case of having only dispersive coupling where we assume $g_{\rm diss}=0$ we get
\begin{eqnarray}
X^\varphi_{\rm out}(\omega)\vert_{\rm disp}&=&\chi_{\rm m}(\omega)(e^{-i\varphi}B(\omega)+e^{i\varphi}B^*(\omega))F_{\rm T}+\chi_{\rm m}(\omega)\zeta A(e^{-i\varphi}B(\omega)+e^{i\varphi}B^*(\omega))rP_{\rm D}(\omega)\nonumber\\ &&+ \vert D(\omega)\vert X_{\rm in}^\theta+\vert E(\omega)\vert X_l^\phi,
\label{x-out-disp}
\end{eqnarray}
where we have used this convention in the Fourier frequency domain that $[a(\omega)]^\dagger=a^\dagger(-\omega)$. For the case of having only dissipative coupling where we assume $g_{\rm disp}=0$ we have
\begin{eqnarray}
X^\varphi_{\rm out}(\omega)\vert_{\rm diss}&=&\chi_{\rm m}(\omega)(e^{-i\varphi}C(\omega)+e^{i\varphi}C^*(\omega))F_{\rm T}+\chi_{\rm m}(\omega)\zeta A(e^{-i\varphi}C(\omega)+e^{i\varphi}C^*(\omega))rP_{\rm D}(\omega)\nonumber\\ &&+ \vert D(\omega)\vert X_{\rm in}^\theta+\vert E(\omega)\vert X_l^\phi,
\label{x-out-diss}
\end{eqnarray} 
where the area of the sensor, $A$, is the area of the disk.

Based on the above mentioned convention, $[a(\omega)]^\dagger=a^\dagger(-\omega)$, the power spectrum of the observable, $X_{\rm out}$, is defined as \cite{BowenMilburn}
\begin{eqnarray}
S_{X_{\rm out}X_{\rm out}}(\omega)&=&\int^\infty_{-\infty}d\omega^\prime\langle X_{\rm out}^\dagger(-\omega)X_{\rm out}(\omega^\prime)\rangle\nonumber\\ &=&\int^\infty_{-\infty}d\omega^\prime\langle (a(\omega)a^\dagger(-\omega))(a^\dagger(-\omega^\prime) a(\omega^\prime))\rangle.
\end{eqnarray}
The power spectrum, $S_{X_{\rm out}X_{\rm out}}$, can be used to calculate the noise equivalent pressure sensitivity. Considering a signal to noise ratio (SNR) equal to unity, the noise power spectrum becomes
\begin{equation}
S_{\rm PP}[\dfrac{{\rm pa}^2}{{\rm Hz}}]=\frac{1}{r^2A^2} (2m\gamma k_{\rm B}T+\frac{1}{N\vert \chi(\omega)\vert^{2}}),
\label{disp_eqn}
\end{equation}
where we define $\chi(\omega)$ as the {\it optomechanical susceptibility}, which depends on the particulars of the coupling regime.
For dispersive coupling it is
  \begin{equation}
\chi(\omega)=\frac{32g_{\rm disp}\Delta\kappa_{\rm{in},0} \chi_{\rm m}(\omega) (\kappa_0-i\omega)}{(4\Delta^2+\kappa_0^2)(4\Delta^2+(\kappa_0-2i\omega)^2)},
\end{equation} 
and for dissipative regime it is
\begin{equation}
\chi(\omega)=\frac{\left [ 2g_{\rm diss}\kappa_{\rm{in},0}(-\kappa_0(\kappa_{\rm in,0}-\kappa_{\rm l})(\kappa_0-2i\omega)+4\Delta^2(\kappa_0+2\kappa_{\rm in,0}-2i\omega) \right ] \chi_{\rm m} (\omega)}{(4\Delta^2+\kappa_0^2)(4\Delta^2+(\kappa_0-2i\omega)^2)}.
\end{equation}  

For the valid regime in this work, $\omega\ll\kappa$, the optomechanical susceptibility reduces to a simpler form of
\begin{equation}
\chi(\omega)=\frac{2g_{\rm i}\kappa_{\rm{in},0} \chi_{\rm m}(\omega)}{(4\Delta^2+\kappa_0^2)^2}\times C^{\rm i},
\end{equation}
where $ i\in \{{\rm disp}, {\rm diss} \}$, $C^{\rm disp}=16\kappa_0\Delta$ and $C^{\rm diss}=-\kappa_0^2(\kappa_{\rm in,0}-\kappa_{\rm l})+4\Delta^2(\kappa_0+2\kappa_{\rm in,0})$.

\section{Sensor response}
In equations (\ref{x-out-disp}) and (\ref{x-out-diss}), the coefficient in front of the external drive force by the applied pressure, $P_d(\omega)$, is the response of the system. In Fig.~\ref{Fig_sup1}, we have plotted the system response versus detuning for both dispersive and dissipative regimes.
   \begin{figure}[!htbp]
 \includegraphics[scale=0.48]{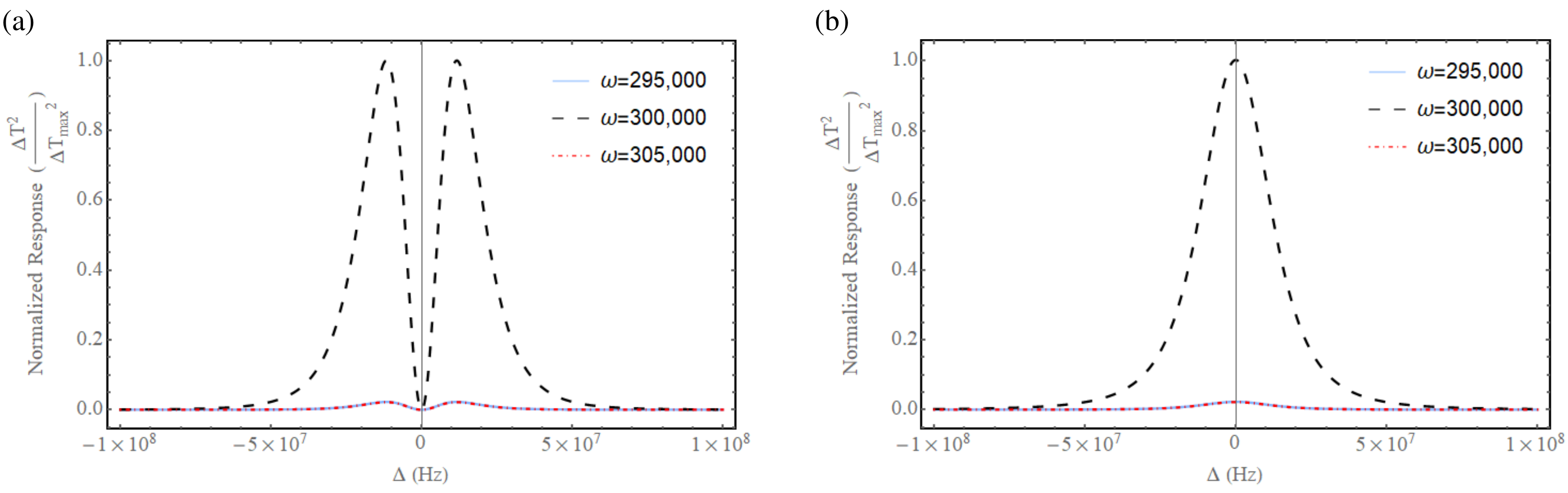}
\caption{{\bf Sensor response versus detuning of the laser form cavity resonance frequency.} Plots are for $\kappa_{\rm l}=4\times 10^7$, $\kappa_{\rm in}=0.5\times 10^6$, and $\omega_{\rm m}=300$ kHz,(a) For the case where the sensor operates in pure dispersive optomechanical regime. (b) for sensor operating in pure dissipative coupling.  } 
\label{Fig_sup1}
    \end{figure}

\section{Utilising a Michelson interferometer to calibrate the piezoelectric sound source}

To calibrate the piezo element (PZT1), we attached to it a light weight silver mirror (M$_1$) which is utilized as a mirror to be displaced in one of the arms of a Michelson interferometer as shown in Fig.~\ref{Interferometer}a. The interferometer is fed by a laser at $\lambda\simeq1555$ nm and the output interference signal is detected on a low noise photodetector as shown in the experiment scheme. We used a secondary PZT element (shown as Phase control PZT in Fig.~\ref{Interferometer}a) with a mirror attached to it in the other interferometer arm to thermally lock its phase using a PID (proportional� integral� derivative) controller.  

\begin{figure}[!htbp]
 \includegraphics[scale=0.455]{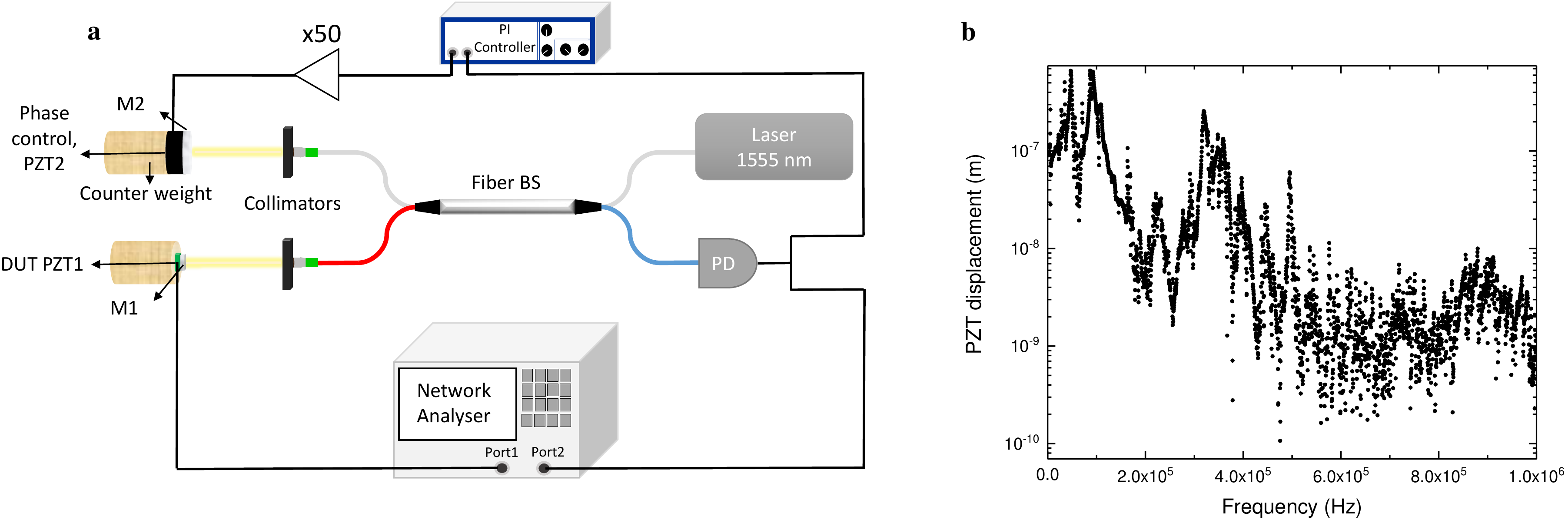}
\caption{{\bf Michelson interferometer to calibrate the piezo-electric sound source.} (a) Interferometry measurement scheme. There are two mirrors M$_1$ and M$_2$ which are respectively attached to PZT1 and PZT2 which are placed in the output arms of the interferometer. PZT1 is the piezo element to be measured and PZT2 is used to lock the phase of the interferometer. PZT1 and PD are respectively connected to ports 1 and 2 of a NA with which we drive the PZ1 and simultaneously measure the signal of the interferometer. (b) The measured displacement of the PZT1 for a drive voltage of 707 mV at frequency spectrum from 1 kHz to 1MHz. } 
\label{Interferometer}
      \end{figure}

To perform the measurement, we used a network analyser with its port 1 driving the PZT1 (DUT) and port 2 receives the signal from the PD. The PI-controller output is connected to PZT2 (phase control) through a voltage amplifier (Falco Sytems WMA-300). The displacement spectrum of PZT1 can be calculated as
\begin{equation}
d(\omega)=\frac{\lambda}{4}\frac{V(\omega_{\rm ref})}{V_{\rm max}}\sqrt{\frac{S_{21}(\omega)}{S_{21}(\omega_{\rm ref})}},
\end{equation}
where $S_{21}(\omega)$ is the off-diagonal network scattering parameter corresponding to the coherent power transfer from port 1 to port 2 at a frequency $\omega$. $\omega_{\rm ref}$ is a calibration reference frequency which was 20 kHz in our measurement. $V(\omega_{\rm ref})$ is the photodetector voltage at $\omega_{\rm ref}$ and $V_{\rm max}$ is the maximum voltage generated by the interference, corresponding to a $\frac{\lambda}{4}$ displacement. Throughout this measurement we always monitored the generated signal not to saturate i.e., the displacement was always $< \frac{\lambda}{4}$ at a given applied voltage to the PZT1. We confirmed that the displacement was a linear function of the applied voltage to PZT1. At frequencies where the displacement was larger than $\frac{\lambda}{4}$, we lowered the voltage in order to avoid saturation. The high and low voltage measurements were then normalised with respect to 707 mV and compiled. The results are shown in Fig.~\ref{Interferometer}b for an applied voltage of 707 mV.\\ \\
The acoustic pressure at the position of the PZT is calculated from the PZT displacement and the air impedance, $\alpha$=413~sPa/m as
\begin{equation}
P_{PZT}(\nu)=\pi \nu d(\nu) \alpha,
\end{equation}
and the acoustic pressure at the position of the sensor is given by
\begin{equation}
P_{sensor}(\nu)=c(\nu) \gamma(\nu)^{-1} P_{PZT}(\nu),
\end{equation}
where c is an attenuation coefficient dependent on the sensor-PZT distance ($L$). We calcaulted this factor based upon on-axis diffraction of a plane wave from the PZT mirror which acts as an apperture through which the sound wave is diffracted. In our measurement the PZT-sensor distance was 10 cm and the appreture size was 7~mm x 7~mm. The results of this calculation as a function of ultrasonic frequency is shown in Fig \ref{attenuation}a. $\gamma(\nu)$ is the atmospheric acoustic attenuation and depends on both $L$ and air acoustic absoprtion coefficient~\cite{air-absorption} which is significant only at high frequencies ($>$100kHz) for small $L$=10~cm. This attenuation factor is shown in Fig \ref{attenuation}b.

\begin{figure}[!htbp]
 \includegraphics[width=1 \textwidth]{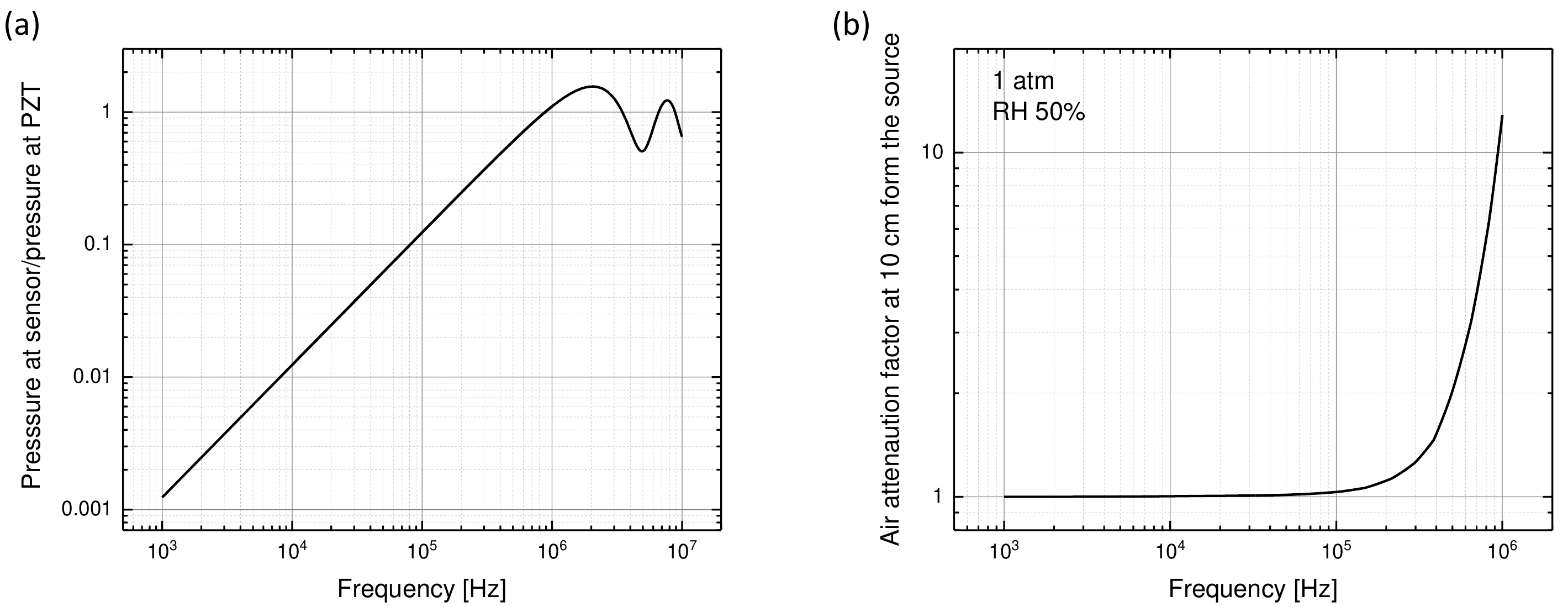}
\caption{ (a) atmospheric attenaution for sensor-PZT distance of 10~cm based on citation \cite{air-absorption}, (b) the ratio of pressure at the position of the sensor to ultrasonic pressure at the PZT calculated from on-axis diffraction for sensor-PZT distance of 10~cm. }
\label{attenuation}
      \end{figure}

\section{Device fabrication process}

Fig.~\ref{figFab} shows the process used to fabricate the ultrasound sensor.

\begin{figure}[h!]
  \includegraphics[scale=0.4]{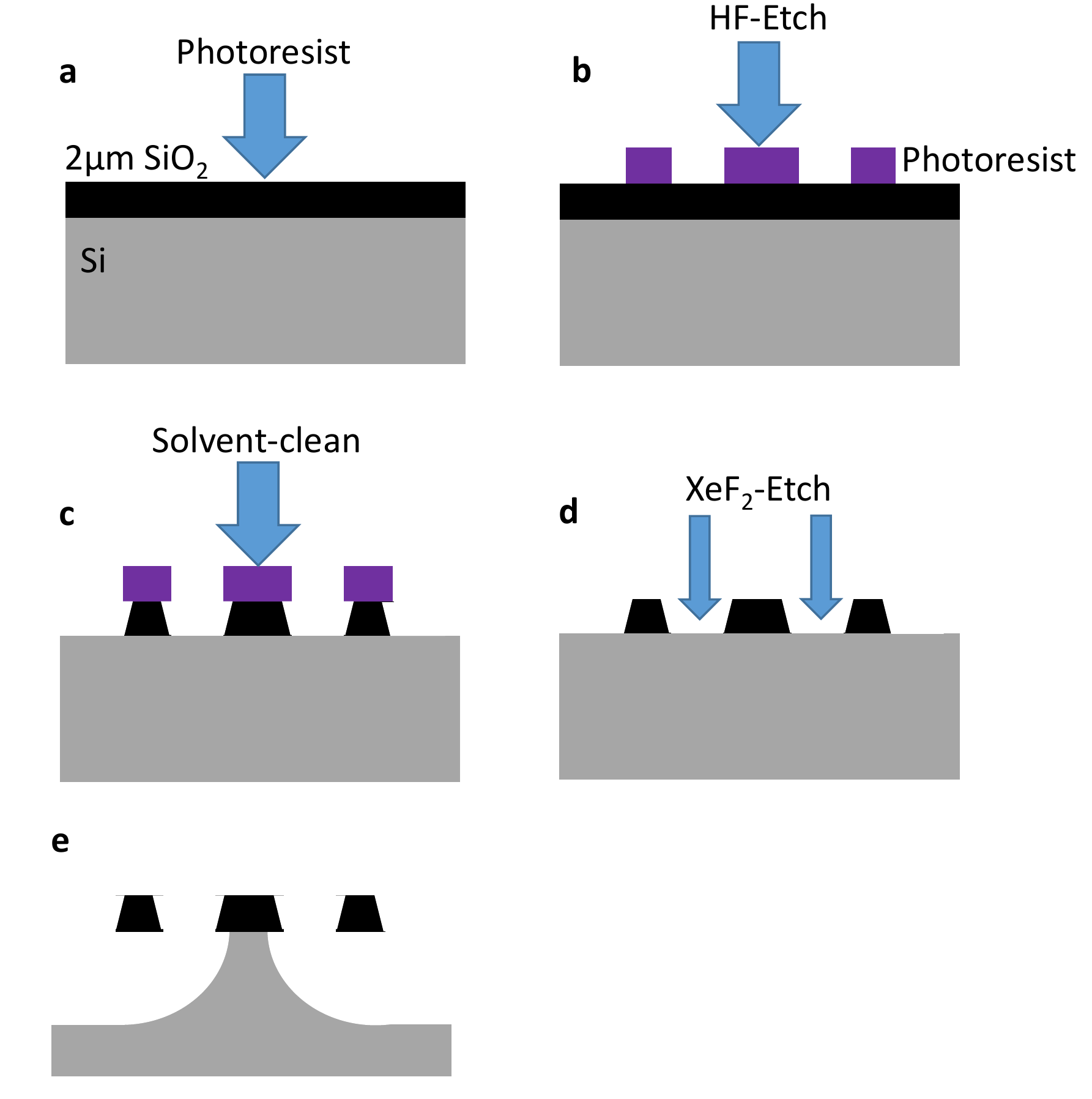}
 
\caption{{\bf Cross-sectional view of the microfabrication process.} a)-c):Starting from a Oxide-coated Silicon wafer, UV-photolithography and HF-wet-etch were performed to define the Silica structure. d) and e): A XeF$_2$ dry-etch isotropically removes silicon and releases the silica-structure.}
 \label{figFab}
\end{figure}

\section{Overlap and effective mass for the second order flapping mode}

In order to compare the measured sensitivity near the second order flapping mode to theoretical predictions, it is necessary to estimate the overlap $\zeta$ between the applied acoustic pressure wave and the mechanical mode spatial profile, as well as the effective mass of the mechanical mode.

The overlap can be calculated as
\begin{equation}
\zeta=\int_A u(\vec{r})\delta p(\vec{r})dA, \label{overlappppp}
\end{equation}
 where the integral is taken over the surface of the resonator; and $u(\vec{r})$ and $\delta p(\vec{r})$ are the mechanical displacement in the direction of the vertically incident pressure wave and the pressure acting on the resonator, respectively. $u(\vec{r})$ is normalized to equal unity at the maximum displacement of the mode and $\delta p(\vec{r})$ is normalized to the acoustic pressure at an antinode of the pressure wave.
 
 The effective mass can be calculated as 
 \begin{equation}
m=t \rho \cdot\int_A |u(\vec{r})|dA,  \label{meee}
\end{equation}
 where $t$ is the thickness of the resonator, $\rho$ is it's density, so that $M = t A \rho$ is the total mass of the resonator.
 
We used COMSOL multiphysics to determine $u(\vec{r})$ for the second order flapping mode.  From Eq.~(\ref{meee}) we then found that the effective mass of the mode was equal to about half the total mass of the resonator, $m \approx 0.5 M \approx 110$~ng.

Considering that the pressure on the sensor, $\delta p$, is due to an incident plane wave,  we found an overlap of approximately $\zeta = 0.14$ from Eq.~(\ref{overlappppp}).

One might expect this overlap to be zero, since the mode exhibits a rotational symmetric tilting motion around the axis of the device with downwards motion at the inner edge of the device at the same time as the outer edge is moving upwards. This counter-motion would lead to a cancellation of the force applied by a plane pressure wave. However, the node of the mode is closer to the outer edge than the center-of-mass of the annulus, leading to a residual center-of-mass motion and a non-vanishing overlap.


\section{Sensitivity and Responsivity}

The experimentally measured pressure sensitivity (black line and blue circles) and the applied pressure (red line) are shown as a function of frequency in Fig \ref{suppsen}. The sensitivity is calculated using Eq.~(2) in the main text and using the measured responsivity and noise floor of the device. The experimentally measured responsivity of the sensor in V/Pa is shown in Fig. \ref{supresp}. As can be observed and might be expected, the response is stronger at low frequencies though, as can be seen in the main text, the noise floor is also increased at low frequencies due to $1/f$ noise. The response also exhibits sharp resonances, as expected for a resonantly enhanced sensor.

\begin{figure}[!htbp]
 \includegraphics[width= 1 \textwidth]{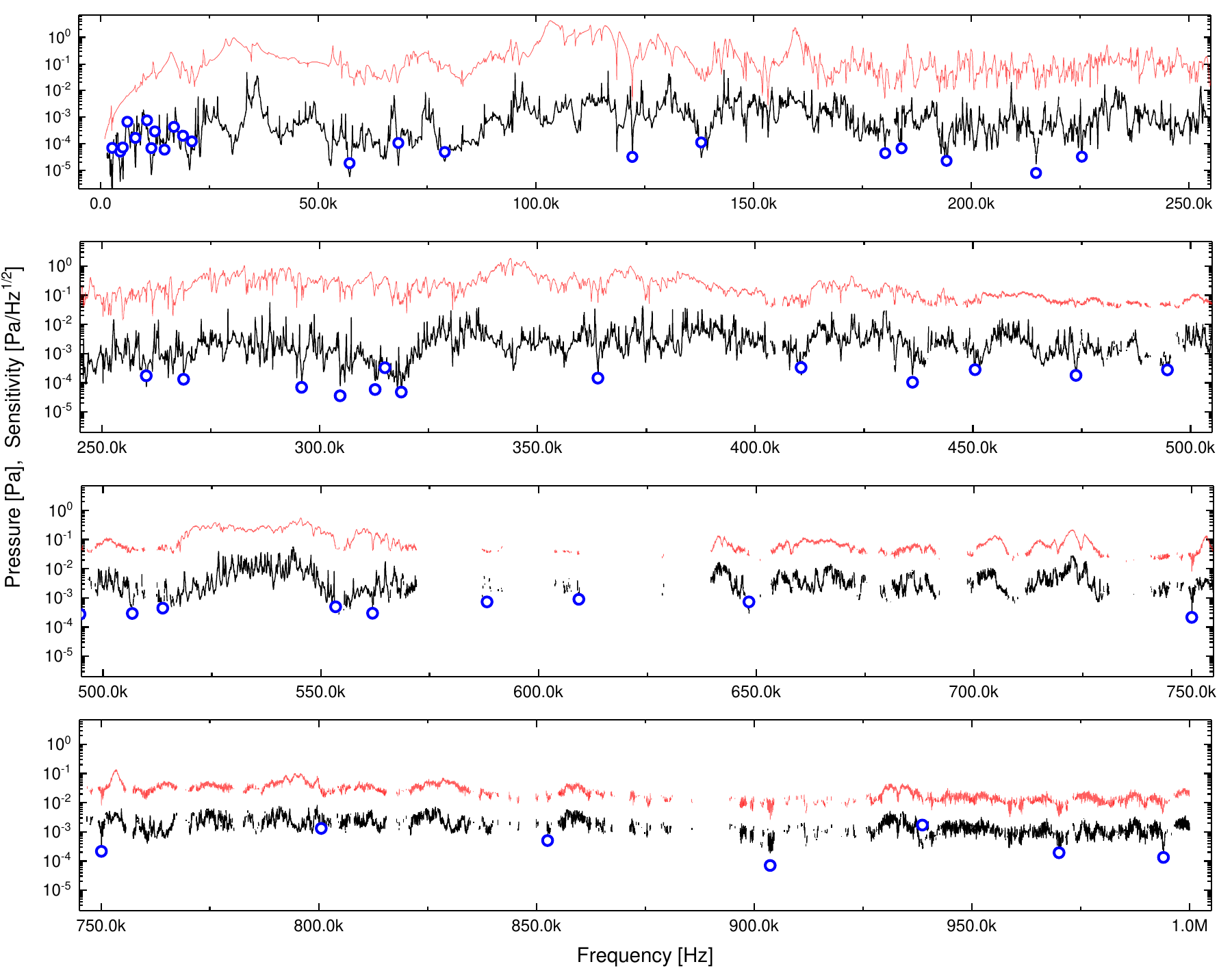}
\caption{{\bf Noise equivalent pressure sensitivity.} Pressure sensitivity as a function of frequency (black lines) and the applied pressure (red lines) as measured using a network analyser. The open symbols are reference points directly measured using a spectrum analyser at certain frequencies for further validation of the network analysis.} 
\label{suppsen}
      \end{figure}

\begin{figure}[!htbp]
 \includegraphics[width=0.7 \textwidth]{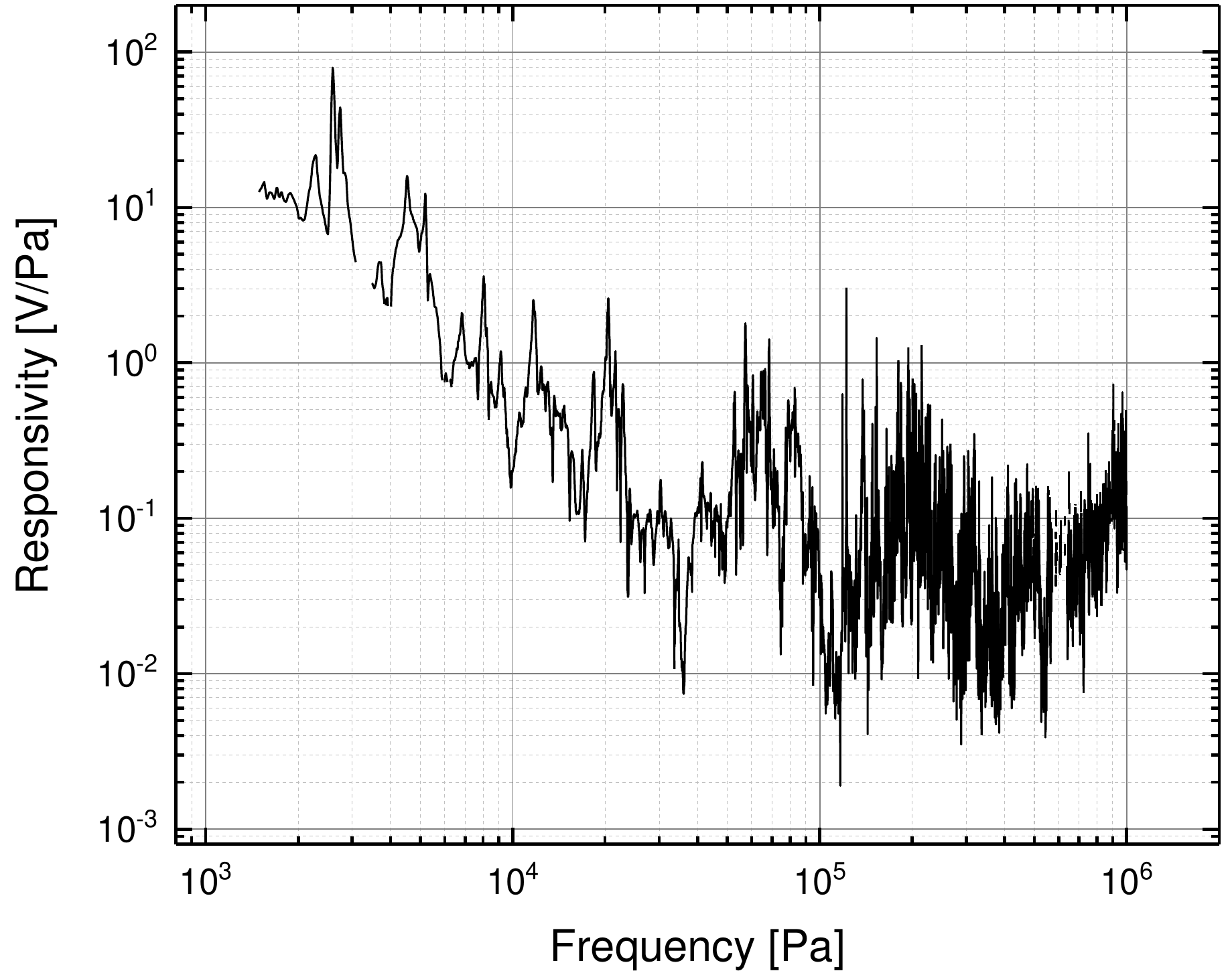}
\caption{Responsivity of the sensor in V/Pa. } 
\label{supresp}
      \end{figure}

\section{Fluidic damping of micromechanical device}

In this section we outline expressions and describe an experiment that allows the fluidic damping due to the interaction of our spoked-disk micromechanical resonator with its gaseous environment to be estimated. We follow reference~\cite{Bao}. There are three common forms of fluidic damping relevant to micromechanical devices: ballistic damping, due to collisions of gas molecules with the surface of the resonator; drag force damping, due to viscousness of the gas and the velocity gradient between the boundary layer near the surface of the resonator and more distant points in the fluid; and squeeze-film damping, due to the change in pressure introduced by motion of the resonator near its substrate. Ballistic damping is generally only significant in high vacuum conditions, and is therefore not considered further here. The other two forms of damping, in general, introduce a combined force that opposes the velocity of the resonator. Following Chapter 3 in reference~\cite{Bao}, and treating the resonator as thin, so that the spatiotemporal eigenmodes ${\bf u}(x,y,z,t)$ which describe the displacement of each small region of the resonator vary only in the plane of the resonator (defined as the $\{x,y\}$ plane here), and do not depend on the $z$-coordinate (i.e. ${\bf u} (x,y,z,t)= {\bf u}(x,y,t)$), this force can be written as $F = - \mu l \dot {\bf u} (x,y,t) $ where $\mu$ is the coefficient of viscosity of the fluid and $l$ is a geometry-dependent characteristic length-scale  to be determined later. To find the rate of damping due to the gas $\gamma_{\rm gas}$, this force should be compared to the acceleration of the resonator $F_{\rm accel}= m \ddot {\bf u} (x,y,t)$, where $m$ is the effective mass of the mode described by $\bf u$. Considering the acceleration and damping terms in the general equation of motion for harmonic oscillation $m \ddot x + m \gamma \dot x + k x = F_{\rm ext}$, where $k$ is the spring constant and $F_{\rm ext}$ the external force, we see that $\gamma_{\rm gas}$ is given simply by
\begin{equation}
\gamma_{\rm gas} = \mu l/m,
\end{equation}
in angular units.

The power spectral density of the thermal force noise introduced by fluctuation-dissipation to complement this fluidic dissipation is
\begin{equation}
S_{T, \rm gas} = 2 m \gamma_{\rm gas} k_B T = 2 \mu l k_B T, \label{dsfgsdfg}
\end{equation}
where, of course, $k_B$ is the Boltzmann constant, and $T$ is the temperature of the system. Since the incident acoustic wave travels within the gaseous medium, $S_{T, \rm gas}$ presents a fundamental bound on the pressure sensitivity of a micromechanical acoustic sensor of fixed geometry and a particular gaseous medium. 

Given that the thermal fluctuations introduced by the interaction with the gas are independent from those introduced by thermal vibrations of the substrate and any other damping mechanisms intrinsic to the resonator, the total thermal force noise experienced by the resonator is
\begin{equation}
S_T = 2 m (\gamma + \gamma_{\rm gas} ) k_B T = 2 ( m \gamma + \mu l ) k_B T, \label{fgnwsxa}
\end{equation}
where $\gamma$ is the intrinsic decay rate of the resonator, and the total mechanical decay rate $\gamma_m = \gamma + \gamma_{\rm gas}$. Inserting this expression into equation~(\ref{disp_eqn}) we find the minimum detectable pressure
\begin{equation}
P_{\rm min}(\omega)= \sqrt{S_{\rm PP}(\omega)} = \frac{1}{r\zeta A}\sqrt{2 ( m \gamma + \mu l ) k_B T+N^{-1}\vert \chi(\omega)\vert^{-2}}, \label{P_min_eqn}
\end{equation}
as given in the main text, where $r$ is the pressure participation ratio (see main text Fig.~2d), and $\omega$ is the drive frequency of the acoustic wave. Note that, since the sensor and detection system are linear, the effect on inefficiencies in detection are simply to transform the effective intracavity photon number from $N \rightarrow \eta N$ where $\eta$ is the efficiency with which light leaves the optical resonator and is successfully detected at the detector.

\subsection{Experimental characterisation of the gas damping}

In order to determine the contributions to the noise equivalent pressure from intrinsic mechanical dissipation and from fundamental gas damping, we placed the device in a vacuum chamber and swept the pressure from 0.056 ~mbar to atmosphere. We monitored the damping rate of the resonance observed at 315~kHz.  At the lowest measured pressure, the decay rate plateaus to a minimum of $150$~Hz, corresponding to the  intrinsic mechanical dissipation $\gamma$, whereas at atmospheric pressure, the decay rate reaches $1,430$~Hz. The difference between these two values corresponds to a gas damping rate of $ \gamma_{\rm gas}/2\pi=1,260$~Hz. Mechanical resonances for three different pressures, measured using a spectrum analyser, are shown in Fig.~\ref{pressure_variation}.

The contributions to the noise equivalent pressure from intrinsic dissipation and gas damping could potentially also be distinguished by varying the viscosity of the gas in other ways; for instance by changing the constituents or temperature of the gas. Evacuating the sample chamber is particular attractive because it suppresses the gas damping by several orders of magnitude without affecting the intrinsic dissipation, and
therefore allows a direct and accurate measurement of the intrinsic dissipation. By contrast, decreasing the temperature of the enclosure by 100 degrees to $\sim$200~K would only reduce the air viscosity by a factor of two. Similarly, replacing the air with alternative gas such as carbon dioxide, hydrogen, helium or xenon would also only alter the viscosity by a factor of two or less.  A further complication associated with changing the temperature is that the intrinsic mechanical dissipation is also temperature dependent, for instance decreasing by roughly a factor of four with a 100 degree decrease in temperature for devices similar to ours in Ref.~\cite{Arcizet}.

\begin{figure}[h!]
  \includegraphics[scale=0.7]{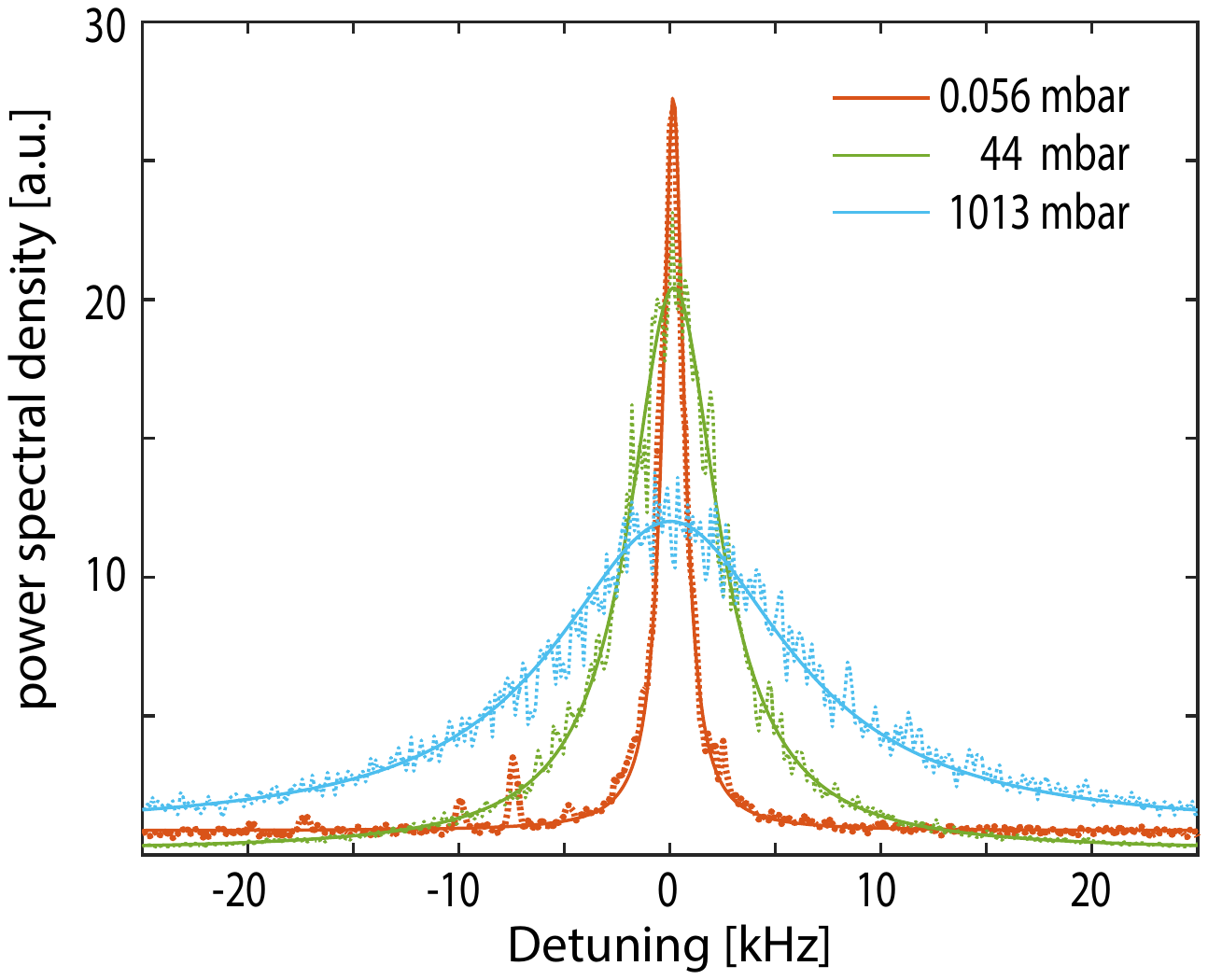}
 
\caption{{\bf Brownian noise spectra of a mechanical resonance with varying static pressure.} Power spectral density versus detuning around the $315$~kHz resonance for different pressures with decay rates of $1430$~Hz, $535$~Hz and $150$ Hz at pressures of $1000$~mbar, $44$~mbar and $0.056$~mbar respectively. Note that the vertical axis of this plot is uncalibrated, and varied for measurements at different pressures due to drifts in the experimental apparatus, including the taper-microdisk separation and the optical polarisation.}
 \label{pressure_variation}
\end{figure}

\subsection{Determining the characteristic length-scale}

From the experimentally observed gas damping of $ \gamma_{\rm gas}/2\pi=1,260$~Hz we can directly determine a viscous length-scale  $l= 2\pi\cdot 8.1$~mm. The length scale has two components -- one from drag damping ($l_{\rm drag}$), and one from squeeze film damping ($l_{\rm squeeze}$). While the viscous drag damping can be reliably calculated for our geometry, as shown in the following paragraphs, the viscous squeeze film damping depends sensitively on the height profile of the underlying substrate. In our case, large  height variations arise in the silicon substrate below the sensor due to our fabrication process, as can be seen in Fig.~2a of the main text. This makes the exact determination of the viscous squeeze film damping extremely challenging, and beyond the scope of this work. We instead infer the characteristic length scale for squeeze film damping  from the experimentally extracted total gas damping and the calculated viscous damping.

\subsubsection{Determining the viscous drag length scale}

Chapter 3 of Ref.~\cite{Bao} gives the drag force for several geometries, with the general form
\begin{equation}
F_{\rm drag} = - 6 \pi \xi \mu \sqrt{A} \, \dot {\bf u} (x,y,t),
\label{eqdrag}
\end{equation}
where, again, $\mu$ is the coefficient of viscosity of the fluid, which for air at room temperature is around $\mu=1.8 \times 10^{-5}$~kg/ms, $A$ is the surface area of the top surface of the resonator and $\xi$ is a dimensionless geometry dependent coefficient which is generally on the order of unity. For instance, for a free sphere, a vertically moving disk ($z$ direction), and a horizontally moving disk, it is given by $\xi = \{1, 0.85, 0.567\}$, respectively.

From Eq.~(\ref{eqdrag}), we find 
\begin{eqnarray}
l_{\rm drag} &=& {6 \pi \xi \sqrt{A}} =  {6 \pi^{3/2} \xi \sqrt{R^2-r^2}} \approx  3 \pi \left (  2 \xi  \sqrt{A} \right ), \label{dfgsdfgas}\\
\gamma_{\rm drag} &=& \frac{\mu l}{m} = \frac{3 \pi \mu}{m} \left (  2 \xi  \sqrt{A} \right ). \label{dfghh}\
\end{eqnarray}

 It should be noted that in these expressions $m$ is the effective mass of the mechanical eigenmode, which we determine to be $m\sim110$~ng~$\sim M/2$ from finite element modelling, while the drag damping calculations assume a uniform vertical motion of the disk - that is, it does not account for any structure in the mechanical modeshape. 

Calculations that fully account for the modeshape dependence of gas damping are beyond the scope of this work. However, it can be approximately accounted for via a simple thought experiment. We imagine that the component of the annular disk is such that the surface of the disk within a radius $r'$ of its axis is perfectly stationary, while the component between $r'$ and the major radius of the disk $R$ moves uniformly. That is, the mechanical eigenmode is a step-function in the radial direction, with no motion at radii less than $r'$. In this case, the effective mass of the resonator is equal to 
\begin{equation}
m = M \times \frac{A'}{A} = M \times \left ( \frac{R^2-r'^2}{R^2-r^2} \right ),
\end{equation}
where $A'$ is the area of the moving component of the annular disk and $M$ is the total mass of the annular disk.

Since the moving component of the annular disk is also an annular disk with the same major radius $R$, but minor radius increased to $r'$, Eqs.~(\ref{dfgsdfgas})~and~(\ref{dfghh}) can be applied to approximate the drag damping and its characteristic length scale, but with the replacement $A \rightarrow A'$.  We then arrive finally at
\begin{eqnarray}
l = l_{\rm drag}\approx 6 \pi  \xi  \sqrt{\frac{A m}{M}},\label{drag2}\\
 \gamma_{\rm drag}  = \frac{\mu l_{\rm drag}}{m} \approx \frac{6 \pi \mu}{m}  \xi  \sqrt{\frac{Am}{M}}  . \label{dfghh2}
\end{eqnarray}
The relevant parameters of our device are $\mu=1.8 \times 10^{-5}$~kg/m s, $R = 148$~$\mu$m, $r = 82$~$\mu$m, a mass of around $M \sim \rho t A  = 230$~ng, and an effective mass of $m \sim M/2$, where $\rho = 2650$~kg/m$^3$ is the density of silica, $t=1.8 \mu$m the thickness of the resonator, and $A$ its area. We  choose $\xi = 0.85$, consistent with expectations for a vertically moving disk.

In order to state a sensitivity in units of Pa/$\sqrt{\rm Hz}$, where Hz is the inverse of the actual measurement time, we include an extra scaling factor of $2\pi$ to convert from radial units that are used in the derivations above: $l\rightarrow l/2\pi=m\gamma\big/2\pi\mu$. We then find
\begin{eqnarray}
l_{\rm drag} &\sim&  0.4~{\rm mm},\\
\gamma_{\rm drag}/2\pi &\sim&  62~{\rm Hz}.
\end{eqnarray}
From this, we infer an approximate squeeze film damping characteristic length scale of  $l_{\rm squeeze}= l-l_{\rm drag}=7.7$ mm $\gg l_{\rm drag}$, suggesting that squeeze film damping is dominant, and thus the total gas damping could be significantly reduced by increasing the distance between the sensor and the substrate. 

\subsubsection{Squeeze film damping in the presence of a flat substrate}

In this Section we will derive an approximate expression for the squeeze film damping above a flat substrate. While, due to the large height variations of the substrate across the area of our device, this analysis can not be used to determine $l_{\rm squeeze}$, we can use it to assess how an optimized sensor geometry could be designed.

Ref.~\cite{Bao} calculates the squeeze film force for a vertically moving annular disk with major and minor radii of $R$ and $r$, respectively, to be
\begin{equation}
F_{\rm squeeze} = - \frac{3 \pi \mu R^4 G(\beta)}{2  h^3} \, \dot {\bf u} (x,y,t),
\end{equation}
where $h$ is the separation of the bottom of the resonator from the substrate, $\beta=r/R$ and 
\begin{equation}
G(\beta) = 1 - \beta^4 + \frac{(1-\beta^2)^2}{\ln \beta}.
\end{equation}

Using these expressions, we find directly that
\begin{equation}
l_{\rm squeeze} = \frac{3 \pi R^4 G(\beta)}{2 h^3}.
\end{equation}

To account, roughly, for the eigenmode shape of the mechanical mode of the sensor, following the approach taken in the previous section,
we can redefine a modified $\beta$ as
\begin{equation}
\beta' = \frac{r'}{R} = \sqrt{1 - \frac{A}{\pi R^2} \frac{m}{M}} = \sqrt{1- \left (1 - \beta^2 \right ) \frac{m}{M}},
\end{equation}
where we have used the fact that $A/\pi R^2 = 1 - \beta^2$. We then find
\begin{eqnarray}
 l_{\rm squeeze} \approx \frac{3 \pi  R^4 G(\beta')}{2 h^3},\label{dfgsdfgas2}\\
 \gamma_{\rm squeeze}  \approx\frac{3 \pi \mu R^4 G(\beta')}{2 m h^3} . \label{dfghh2}
\end{eqnarray}

 The relations derived in this Section for the characteristic length scales and damping rates for air-drag damping and squeeze-film damping, combined with modelling of the intrinsic mechanical damping, such as that performed in Ref.~\cite{Kippenberg spoked toroids}, allow approximate prediction of the thermomechanical noise limited sensitivity of a general spoked-microdisk cavity optomechanical acoustic sensor.

\subsubsection{Designing the device to reach optimum sensitivity}

As given in Eq.~(\ref{dsfgsdfg}), the thermal force noise introduced by fluidic dissipation, and therefore the gas-damping limited pressure sensitivity, only depends on the Boltzmann constant, the temperature, the gas viscosity and the characteristic length-scale. Therefore, the characteristic length-scale is the only device-dependent parameter which can be engineered to optimise the sensitivity in this limit. The right-hand-side of Eq.~(\ref{drag2}) represents the characteristic length-scale for air-drag damping. As can be seen, apart from the geometric-factor $\xi$ this depends only on the area of the resonator which is fixed for a given desired spatial resolution and the ratio of mass to effective mass. The air-drag damping limited sensitivity is therefore also fixed for given desired spatial resolution and device geometry. On the other hand, the squeeze-film damping characteristic length-scale (right hand side of Eq.~(\ref{dfgsdfgas2})) depends on the height of the device above the substrate $h$, decreasing with increasing height. The two characteristic length-scales are equal for $h = (R^{4}G(\beta')/(4 \xi \sqrt{Am/M}))^{1/3}$. Taking the rough approximation that $A \sim R^2$, $m=M$, and $\beta \sim 0$ so that the hole in the annular disk is small relative to its diameter, this becomes $h \sim R$. We therefore see that squeeze-film damping can be expected to significantly degrade the pressure sensitivity if  the height of the device is small compared to its radius. This is the case for our existing devices as can be seen from Fig.~2a in the main text, consistent with our observation above that squeeze film damping dominates drag damping for our devices. Given the $1/h^3$ scaling in Eq.~(\ref{dfgsdfgas2}), it is clear that by developing a modified fabrication process that allow our devices to be suspended further from the substrate, the squeeze-film damping could be greatly suppressed.

It is interesting to also observe that $l$, and therefore the thermal force noise from gas damping, is independent of the thickness of the device. On the other hand, as can be seen in Eq.~(\ref{fgnwsxa}), the force noise due to intrinsic damping into the substrate increases linearly with thickness, through the increase this causes to the device mass.
Consequently, as the resonator becomes increasingly thin, and the intrinsic thermal force noise decreases, the noise introduced by gas-damping will become dominate (as is already the case for our devices). For sufficiently good optical measurement and a sufficient height above the substrate, this would allow the sensor to operate at the air-drag damping force noise floor. Indeed, for a sufficiently thin device, it may be possible to achieve an air-drag damping-limited noise floor even without the presence of spokes to isolate the device from substrate thermal noise. In this case, the active sensing would be increased by around 40\% improving the sensitivity by a similar margin.
}

\section{Estimation of sensitivity of trace gas sensing by photo-acoustic spectroscopy}
Photo-acoustic gas spectroscopy is based upon sensing the acoustic waves generated by gas moleucles due the light absoprtion. Excitation light is properly chosen to be on resonance with one of the spectral lines of the gas molecules. Absorption of light in the gas produces local heating in the sample which results in local pressure increase. If the excitation light is pulsed or a modulated continuous-wave (CW), the generated heat in the gas will result in generation of acoustic waves at the modulation frequency. Photo-acoustic gas sensing is based on measuring the generated acoustic pressure to measure the gas absorption and so the concentration of the sample gas. The optomechanical sensor has high sensitivity together with microscale area. Hence, it offers the possibility to image gas concentrations with high resolution. Here we consider one example, the possibility to measure the CO$_2$ exchange of photosynthetic cells. 

We can estimate the lowest gas concentration which can be measured by the optomechanical acoustic sensor in vicinity of a photosynthetic sample such as a plant leaf~\cite{Ho}. For this. we need to connect the minimum detectable pressure by the opto-mechanical microphone to the optical absorption coefficient in order to calculate the minimum of gas concentration which can be measured. We consider a microscale photosynthetic sample which exchanges CO$_2$ with its environment. We place the acoustic sensor at a distance $r$ from the sample and shine a pulsed laser through the gas in the vicinity of the sample. We choose the spectral line of CO$_2$ at $\lambda=4,329.93$~nm ($k=2311.105~{\rm cm}^{-1}$) having line intensity of $S=4.7\times10^{-19}~\frac{{\rm cm}^{-1}}{{\rm molec.cm}^{-2}}$. 

For the remainder of the analysis we follow~\cite{Tam} to find the relation between absorption coefficient of the gas and the generated photoacoutic pressure. As in~\cite{Tam}, we consider a thin optical medium (low absorption) for which $\alpha l\ll1$ where $\alpha$ is the optical absorption coefficient and $l$ is irradiation length or length of the photoacoustic source. We further assume that the sound wave can exit the irradiated zone within the pulse duration so that $R_{\rm s}<v\tau_{\rm L}$, where $R_{\rm s}$ is the radius of the laser beam, $v$ is the speed of sound and $\tau_{\rm L}$ is the laser pulse duration. Therefore, the effective source radius is $R=v\tau_{\rm L}$. The second assumption can therefore be rewritten as $R_{\rm s}<R$, i.e. that the radius of the laser beam should be smaller than the source radius. Moreover, the source volume can be written as $V=\pi R^2l$. The coefficient of expansion of air, $\beta$ is
\begin{equation}
\beta=\frac{\Delta V}{V\Delta T},
\label{beta}
\end{equation} 
where $\Delta V=\pi (R+\Delta R)^2l$ is the initial expansion of the source volume after the laser beam, and $\Delta T$, the rise in the temperature after a pulse, is~\cite{Tam}
\begin{equation}
\Delta T=\frac{E\alpha l}{\rho V C_p},
\label{DT}
\end{equation} 
in which $E$ is the energy of the laser pulse, $\rho$ is the density and $C_p$ is the heat capacity of air. Therefore, $\Delta R$, the initial expansion of the source radius becomes~\cite{Tam}
\begin{equation}
\Delta R=\frac{\beta E\alpha}{2\pi\rho C_p v\tau_{\rm L}}.
\end{equation} 
The peak displacement, $U_s(r)$ at distance $r$ from the photoacoustic source varies as $\dfrac{1}{r}$ for spherical sound waves. Hence,
\begin{equation}
U_s(r)=\Delta R(\frac{R}{r})=\frac{\beta E\alpha}{2\pi\rho C_p r}.
\label{U_s}
\end{equation}
the peak acoustic pressure is~\cite{Tam}
\begin{equation}
P_{\rm peak}(r)\approx\frac{v\rho U_s(r)}{\tau_{\rm L}}.
\label{P_s}
\end{equation}
Equations~(\ref{U_s}) and (\ref{P_s}) result in
\begin{equation}
	\alpha\approx\frac{2\pi C_p \tau_{\rm L}r}{v\beta E}P_{\rm peak}(r).
	\label{abs}
\end{equation}
To calculate the minimum detectable concentration we first  need to relate the peak acoustic pressure to the effective acoustic pressure driving the mechanical mode over a period of the mechanical motion. The conversion factor can be estimated as the ratio of the laser pulse duration and the mechanical period over which the pressure is being applied. We have
\begin{equation}
P_{\rm eff}(r)=P_{\rm peak}(r)\tau_{\rm L}\frac{\omega_{\rm m}}{2\pi},
\label{p-eff}
\end{equation}
in which $\omega_{\rm m}$ is the mechanical frequency. 
Moreover, the absorption coefficient is proportional to gas concentration as
\begin{equation}
\alpha=\frac{cS}{2\gamma_G},
\label{concentarion}
\end{equation}
in which $S=4.7\times10^{-19}~\frac{{\rm cm}^{-1}}{{\rm molec.cm}^{-2}}$ is line intensiy of ${\rm CO_2}$, $\gamma_G=0.06~{\rm cm}^{-1}$ is gas linewidth of and $c$ is the number density of gas molecules. Using equations (\ref{abs}), (\ref{p-eff}) and (\ref{concentarion}) we can write 
\begin{equation}
c_{\rm min}\approx\frac{8\pi^2 \gamma_GC_p r}{v\beta ES\omega_{\rm m}}P_{\rm eff-min},
\label{cmin}
\end{equation}
where $P_{\rm eff-min}$ is the minimum pressure that can be detected by the optomechanical transducer and $c_{\rm min}$ is the minimum detectable gas molecules number density.
At room temperature, $T=300$ K, $\beta=0.0034~\frac{1}{\rm K}$ and $C_p=1.005~\frac{{\rm kJ}}{{\rm kg.K}}$. We further assume a pulsed laser having a pulse energy of $E=1~\mu{\rm J}$ and a pulse duration of $\tau_{\rm L}=1~\mu$s. This pulse  duration is short enough to satisfy the condition of $R_{\rm s}<v\tau_{\rm L}$ for a typical laser beam radius of $R_{\rm s}=50~\mu $m but also long enough to avoid thermal diffusion during the exposure. If we choose the acoustic frequency of $\nu_0=318$~kHz at which the optomechanical sensor can detect acoustic pressures as small as $P_{\rm min}=84~\mu {\rm Pa}$, using equation (\ref{cmin}) at $r=100~\mu$m, we get $c_{\rm min}=3.5\times10^{11}~\frac{\rm molec}{{\rm cm}^3}$ which is equal to 12.5 ppb. The concentration of CO$_2$ around leaf cells investigated in~\cite{Ho} is of the order of 100 ppm. Therefore, our sensor can be expected to be sensitive enough to measure CO$_2$ exchange of photosynthetic cells with a high resolution.

\section{Measurement of the acoustic waves generated by the nanoscale vibrations of cells or bacteria  }
Bio-identifications are required in various fields including medicine, food and beverages, water, safety, public health and security. Current procedures of bacteria detection are costly, time-consuming and based on cell culture which require laboratory and microscopic analysis done by a trained person. However, self-contained mobile bio-sensors can simplify fast diagnosis in place even for the some bacteria that can not be cultured in laboratory~\cite{Ahmed}.

There are recent experiments on bacteria, yeast and plant cell samples in liquid and soil which show bacteria and yeast produce vibrations with displacement amplitudes of 1-100 nm and plant cells produce vibrations with displacement amplitude of 1-30 nm~\cite{Kasas,Lissandrello,Song,Longo}. These experiments are performed for cell concentrations of 108 and 107 CFU/ml which respectively include 96 and 27 bacteria per sample~\cite{Lissandrello}. 
\begin{figure}
	\includegraphics[scale=0.9]{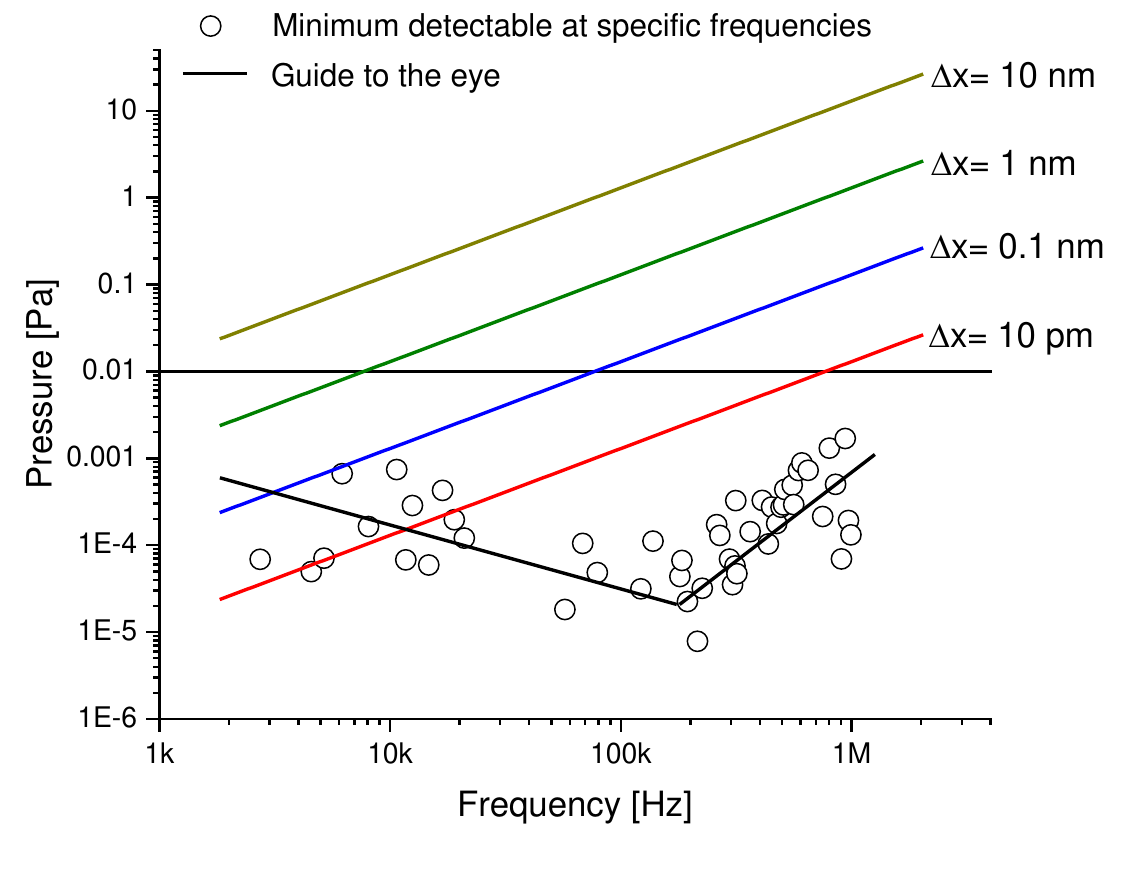}
	\caption{{\bf Opto-mechanical acoustic sensor performance for cell vibration detection.} The circles are experimental sensitivities (at a bandwidth of 1 Hz) for our sensor with given size. Our sensor is scalable and it can be fabricated slightly larger or smaller thereby changing its sensitivity at different frequency. However, more or less the solid black line shows a guide to the eye for expected sensitivity of such sensor we have. The color solid lines show the amount of pressure generated by the cell versus frequency for different displacements. }
	\label{bacteria-detection}
\end{figure}

Our opto-mechanical acoustic sensor can be used to study micro-organisms through detecting these vibrations and hence there is no need to grow cells in an especial probing medium. We can consider a very thin layer of bacteria or cell in a liquid which is coated on a silicon or glass substrate and hold our acostic sensor very close to the sample such that we can ignore the air attenuation of the acoustic waves generated by cells. In order to estimate if our sensor is sensitive enough to probe such vibrations, we need to estimate the pressure produced by these micro-orgnisms vibrations and compare it with the minimum presure that our sensor can detect at a given frequency. The pressure generated by the vibrational displacements can be calculated as
\begin{equation}
P=\pi \nu_0 Z_{\rm air}d,
\end{equation}
in which $\nu_0$ is the vibration frequency, $Z_{\rm air}=$413 Pa.s/m is air impedance and $d$ is the displacement. For $\nu_0=10~{\rm kHz}$ and $d=$1~nm we have $P=1.3\times10^{-2}$~Pa which suggests that our acoustic sensor should be able to quite easily detect cell vibrations.

Fig.~\ref{bacteria-detection} shows how the opto-mechanical sensor performs for cells vibrating at other frequencies and with smaller displacement amplitudes. As can be seen, the sensor should be applicable to sensitively detect small cellular vibrations at frequencies across the full range over which it has been calibrated. We note, further, that the broadband sensitivity of better than 10~mPa/$\sqrt{\rm Hz}$ is sufficient to monitor cellular vibrations across the full continuous frequency range. One technical consideration is that, as with other resonant sensors and as discussed earlier in the Supplement, the responsively of  the sensor fluctuates significantly over the measurement band. These fluctuations are static in time, and therefore could be compensated for in post-processing to produce an accurate spectrum of cellular vibrations. Alternatively, as discussed in the main text and in more detail later in the Supplement,  optomechanical cooling techniques could be employed to broaden the mechanical resonances and flatten the response without degrading singnal-to-noise.

Our opto-mechanical acoustic sensor can also offer some other advantages for cell detection such as using this sensor one does not need an agent as in~\cite{Kasas,Lissandrello,Song,Longo} since the sensor can measure cells vibration without need to attach them to the sensor. Moreover, we can detect the magnitude and frequency of the vibrations and scan over xy coordinates to map the vibrational pattern. This enables studying and investigating different bacteria in a sample. This sensing enables the experimenter to measure spectral density of the acoustic waves generated by the cell which may help to understand the difference between different types of cells (maybe cancer detection) or many other cell biology investigation such as probing fast phenomena happening on the cell wall or inside it. The cell wall can also behave like a membrane transferring internal oscillations to the air and finally to the sensor. This suggest that our sensor has significant potential in developing micromechanical sensors for micro-organisms.

\section{Optomechanical  cooling}

As discussed in the main text, a range of techniques have been developed in the quantum optomechanics community to cool the motion of mechanical resonators (see e.g. \cite{Chan,Teufel,Schliesser,Bowen,Harris,Kim2017}).  These cooling processes also, by necessity broaden the mechanical resonances. This broadening can be used to flatten the resonant response of the sensor. Unlike other methods to broaden mechanical resonance, for example, by introducing additional damping, however, since the ideal laser acts as a zero temperature bath~\cite{BowenMilburn}, these quantum optomechanical cooling techniques do not add additional thermal noise. Therefore, they can be used to flatten the response of an optomechanical system without the usual cost of additional noise. The fractional broadening of the mechanical resonances which is possible can be quantified using a single parameter, the optomechanical cooperativity
\begin{equation}
C = \frac{4 g_0^2 N}{\kappa \gamma},
\end{equation}
where $g_0$ is the vacuum optomechanical coupling rate, $N$ is the number of intracavity photons, $\kappa$ is the optical decay rate, and $\gamma$ is the mechanical resonance linewidth. Optomechanical cooperativities in the range of $10^3$ to $10^6$ can generally be readily achieved~\cite{Aspel}.  Application of, for example, feedback cooling~\cite{Bowen,Harris}, can broaden the mechanical linewidth by as much as a factor of $C$. So, for example, a mechanical resonance at 500~kHz with a quality factor of 1,000 (and therefore linewidth of 500~Hz) could to broadened  to give a near-flat response.


\newpage
\begin{references}

\bibitem{Purdy2017} Purdy T.P., Grutter, K.E., Srinivasan, K., and Taylor, J.M. Quantum correlations from a room-temperature optomechanical cavity. {\it Science}, {\bf 356,} 1265-1268 (2017). 
\bibitem{PainterAccelerometer}Krause, A. G., Winger, M., Blasius, T. D., Lin, Q. \& Painter, O. A high-resolution microchip optomechanical accelerometer. {\it Nat. Photon.} {\bf 6,} 768-772, (2012).
\bibitem{WBmagnetometer}Forstner, S. {\it et al.}  Cavity optomechanical magnetometer. {\it Phys. Rev. Lett.}, {\bf 108,} 120801 (2012).
\bibitem{KippenbergForce}Gavartin, E., Verlot, P. \& Kippenberg, T. J. A hybrid on-chip optomechanical transducer for ultrasensitive force measurements. {\it Nat. Nanotechnol.} {\bf 7,} 509-514 (2012).
\bibitem{Basiri-optical-sensor}Basiri-Esfahani, S., Myers, C. R., Armin, A., Combes, J., \& Milburn, G. J. Integrated quantum photonic sensor based on Hong-Ou-Mandel interference. {\it Opt. Express}, {\bf 23,} 16008-16023 (2015).
\bibitem{PainterPhC}Eichenfield, M., Chan, J.,  Camacho, R. M., Vahala, K. J. \& Painter, O. Optomechanical crystals. {\it Nature}, {\bf 462,} 78 (2009).
\bibitem{BowenMilburn}Bowen, W.P. \& Milburn, G.J. {\it Quantum Optomechanics} (CRC Press, 2015).
\bibitem{Schliesser2008} Schliesser, A., Anetsberger, G., Riviere, R., Arcizet, O. \& Kippenberg, T.J. High-sensitivity monitoring of micromechanical vibration using optical whispering gallery mode resonators. New Journal of Physics, {\bf 10,} 095015 (2008).
\bibitem{gravit wave}Abbott, B. P. et al. GW151226: Observation of gravitational waves from a 22-solar-mass binary black hole coalescence. {\it Phys. Rev. Lett.} {\bf 116,} 241103 (2016).
\bibitem{Lu2016} Yu, W., Jiang, W.C., Lin, Q., \& Lu, T. Cavity optomechanical spring sensing of single molecules. {\it Nature Communications}, {\bf 7,} 12311 (2016). 
\bibitem{Wu}Wu, M. et al. Dissipative and dispersive optomechanics in a nanocavity torque sensor. {\it Phys. Rev. X}, {\bf 4,} 021052 (2014).
\bibitem{WBmagnetometer2}Forstner, S. et al. Ultrasensitive optomechanical magnetometry. {\it Advanced Materials}, {\bf 26,} 6348-6353 (2012).
\bibitem{Kim2017} Kim, P.H., Hauer, B.D., Clark, T.J., Fani Sani, F., Freeman, M.R., Davis, J.P. Magnetic actuation and feedback cooling of a cavity optomechanical torque sensor. {\it Nature Communications} {\bf 8} 1355 (2017).
\bibitem{Teufel}Teufel, J. D. et al. Sideband cooling of micromechanical motion to the quantum ground state. {\it Nature}, {\bf 475,} 359 (2011).
\bibitem{Chan}Chan, J. et al. Laser cooling of a nanomechanical oscillator into its quantum ground state. {\it Nature}, {\bf 478,} 89 (2011).
\bibitem{Schliesser}Schliesser, A. et al. Resolved-sideband cooling of a micromechanical oscillator. {\it Nat. Phys.} {\bf 4,} 415 (2008).
\bibitem{groblacher}Riedinger, R. et al. Non-classical correlations between single photons and phonons from a mechanical oscillator. {\it Nature}, {\bf 530,} 313 (2016).
\bibitem{Schwab}Wollman, E. E. et al. Quantum squeezing of motion in a mechanical resonator. {\it Science}, {\bf 349,} 952-955 (2015).
\bibitem{Andrews2014} Andrews, R.W. et al. Bidirectional and efficient conversion between microwave and optical light. {\it Nat. Phys.} {\bf 10,} 321 (2014).
\bibitem{Chu2017} Chu, Y. et al. Quantum acoustics with superconducting quibits. {\it Science}, eaao1511 (2017).
\bibitem{Basiri-Esfahani}Basiri-Esfahani, S., Akram, U. \& Milburn, G. J. Phonon number measurements using single photon opto-mechanics. {\it New J. Phys.} {\bf 14,} 085017 (2012).
\bibitem{Dong}Wu, H. et al. Beat frequency quartz-enhanced photoacoustic spectroscopy for fast and calibration-free continuous trace-gas monitoring, {\it Nat. Commun.} {\bf 8,} 15331 (2017).
\bibitem{Fischer}Fischer, B. Optical microphone hears ultrasound. {\it Nat. Photon.},  {\bf 10,} 356-358 (2016).
\bibitem{nanofiber}Lang, C., Fang, J., Shao, H., Ding, X. \& Lin, T. High-sensitivity acoustic sensors from nanofibre webs. {\it Nat. Commun.} {\bf 7}, (2016).
\bibitem{resistivity}Scheeper, P. R., van der Donk, A. G. H., Olthuis, W. \& Bergveld, P. A review of silicon microphones. {\it Sensor. Actuat. A-Phys.} {\bf 44}, 1–11 (1994).
\bibitem{capacitive}Chan, C. K., Lai, W.C., Wu, M., Wang, M.Y. \& Fang, W. Design and implementation of a capacitive-type microphone with rigid diaphragm and flexible spring using the two poly silicon micromachining processes. {\it IEEE Sens. J.} {\bf 11,} 2365-2371 (2011).
\bibitem{Miller2016} Miller, J.M., Ultrasound resolution beats the diffraction limit. {\it Phys. Today}, {\bf 69,} 14 (2016).
\bibitem{Ballantine}Ballantine Jr, D. S. et al. {\it Acoustic Wave Sensors: Theory, Design and Physico-Chemical Applications} (Elsevier, 1996).
\bibitem{Wissmeyer} Wissmeyer, G., Pleitez, M. A., Rosenthal, A. \& Ntziachristos. Looking at sound: optoacoustics with all-optical ultrasound detection. Light: Science \& Applications {\bf 7} 53 (2018).
\bibitem{Preisser}Preisser, S. All-optical highly sensitive akinetic sensor for ultrasound detection and photoacoustic imaging. {\it Biomed. Opt. Express}, {\bf 7,} 4171-4186 (2016).
\bibitem{Guggenheim}Guggenheim, J. A. et al. Ultrasensitive plano-concave optical microresonators for ultrasound sensing. {\it Nat. Photon.} {\bf 11,} 714 (2017).
\bibitem{Kim2017B} Kim, K. H. et al. Air-coupled ultrasound detection using capillary-based optical ring resonators. Scientific Reports {\bf 7} 109 (2017). 
\bibitem{dispersivecoupling}Thompson, J.D. et al. Strong dispersive coupling of a high-finesse cavity to a micromechanical membrane. {\it Nature}, {\bf 452,} 72-75 (2008).
\bibitem{Hammerer}Xuereb, A., Schnabel, R. \& Hammerer, K. Dissipative optomechanics in a Michelson-Sagnac interferometer. {\it Phys. Rev. Lett.} {\bf 107,} 213604 (2011).
\bibitem{Knittel} Knittel, J., Jong, J.H., Gray, M.B., Taylor, M.A. \& Bowen W.P., Ultrasensitive real-time measurement of dissipation and dispersion in a whispering-gallery mode microresonator. {\it Opt. Lett.} {\bf 38,} 1915-1917 (2013).
\bibitem{Li2009} Li, M., Pernice, W.H.P., \& Tang, H.X. Reactive cavity optical force on microdisk-coupled nanomechanical beam waveguides. {\it Phys. Rev. Lett.} {\bf 103,} 223901 (2009).
\bibitem{Beibei} Li, Bei-Bei, Bilek, J., Hoff, U.B., Madsen, L.S., Forstner, S., Prakash, V., Schafermeier, C., Gehring, T., Bowen, W.P., and Andersen, U.L.,  Quantum enhanced optomechanical magnetometry. {\it Optica} {\bf 5,} 850-856 (2018). 
\bibitem{Optical-acoustic}Ma, J. et al. Fiber-optic Fabry-P\'{e}rot acoustic sensor with multilayer graphene diaphragm. {\it IEEE Photon. Technol. Lett.} {\bf 25,} 932-935 (2013).
\bibitem{Winkler} Winkler, A. M., Maslov, K. \& Wang, L. V. Noise-equivalent sensitivity of photoacoustics. {\it J. Biomed. Opt. {\bf 18} 097003 (2013).}  
\bibitem{Kippenberg spoked toroids}Anetsberger, G., Rivi\`{e}re, R., Schliesser, A., Arcizet, O., \& Kippenberg, T. J. Ultralow-dissipation optomechanical resonators on a chip. {\it Nat. Photon.} {\bf 2,} 627 (2008).
\bibitem{Vahala}Jiang, X., Lin, Q., Rosenberg, J., Vahala, K., \& Painter, O. High-Q double-disk microcavities for cavity optomechanics. {\it Opt. Express}, {\bf 17,} 20911-20919 (2009).
\bibitem{Baker2011} Baker, C. et al. Critical optical coupling between a GaAs disk and a nanowaveguide suspended on the chip. {\it Appl. Phys. Lett.} {\bf 99,} 151117 (2011).
\bibitem{Ding2010} Ding, L. et al. High frequency GaAs nano-optomechanical disk resonator. {\it Phys. Rev. Lett.} {\bf 105,} 263903 (2010).
\bibitem{LDR}de Arquer, F. P. G., Armin, A., Meredith, P., \& Sargent, E. H. Solution-processed semiconductors for next-generation photodetectors. {\it Nat. Rev. Mater.}, {\bf 2}, 16100 (2017).
\bibitem{Schmid}Schmid, K., Lutz, P., Tomi\'{c}, T., Mair, E., \& Hirschm\"uller, H. Autonomous Vision‐based Micro Air Vehicle for Indoor and Outdoor Navigation. {\it J. F. Robot.}, {\bf 31,} 537-570 (2014).
\bibitem{PASleaf}Ho, Q. T., Verboven, P., Yin, X., Struik, P. C., \& Nicola\"\i, B. M. A microscale model for combined CO2 diffusion and photosynthesis in leaves. {\it PloS One}, {\bf 7,} e48376 (2012).
\bibitem{Oswald}Oswald, R. et al. HONO emissions from soil bacteria as a major source of atmospheric reactive nitrogen. {\it Science}, {\bf 341,} 1233-1235 (2013).
\bibitem{Longo}Longo, G. et al. Rapid detection of bacterial resistance to antibiotics using AFM cantilevers as nanomechanical sensors. {\it Nat. Nanotechnol.} {\bf 8,} 522 (2013).
\bibitem{Alonso-Sarduy}Alonso-Sarduy, L. et al. Real-time monitoring of protein conformational changes using a nano-mechanical sensor. {\it PloS One}, {\bf 9,} e103674 (2014).
\bibitem{Bowen}Lee, K. H., McRae, T. G., Harris, G. I., Knittel, J., \& Bowen, W. P.  Cooling and control of a cavity optoelectromechanical system. {\it Phys. Rev. Lett.} {\bf 104,} 123604 (2010).
\bibitem{Harris}Harris, G.I., Andersen, U.L., Knittel, J., \& Bowen, W. P.  Feedback-enhanced sensitivity in optomechanics: Surpassing the parametric instability barrier. {\it Phys. Rev. A} {\bf 85,} 061802(R) (2012).
\bibitem{Geraci}Geraci, A. A., Papp, S. B., \& Kitching, J.. Short-range force detection using optically cooled levitated microspheres. {\it Phys. Rev. Lett.} {\bf 105}, 101101 (2010).


\end{references}

\newpage
\begin{references}
	
	\bibitem{Kim}Kim, K. H. et al. Air-coupled ultrasound detection using capillary-based optical ring resonators. {\it Sci. Rep.} {\bf 7,} 109 (2017).
	\bibitem{Kuntzman}Kuntzman, M. L., \& Hall, N. A. A broadband, capacitive, surface-micromachined, omnidirectional microphone with more than 200 kHz bandwidth. {\it J. Acoust. Soc. Am.} {\bf 135,} 3416-3424 (2014).
	\bibitem{Jo}Jo, W., Akkaya, O. C., Solgaard, O., \& Digonnet, M. J. Miniature fiber acoustic sensors using a photonic-crystal membrane. {\it Opt. Fiber Technol.} {\bf 19,} 785-792 (2013).
	\bibitem{Hansen}Hansen, S. T., Ergun, A. S., Liou, W., Auld, B. A., \& Khuri-Yakub, B. T. Wideband micromachined capacitive microphones with radio frequency detection. {\it J. Acoust. Soc. Am.} {\bf 116,} 828-842 (2004).
	\bibitem{Bucaro}Bucaro, J. A., Lagakos, N., Houston, B. H., Jarzynski, J., \& Zalalutdinov, M. Miniature, High performance, low-cost fiber optic microphone. {\it J. Acoust. Soc. Am.}, {\bf 118,} 1406-1413 (2005).
	\bibitem{Jo}Jo, W., Kilic, O., \& Digonnet, M. J. Highly sensitive phase-front-modulation fiber acoustic sensor. {\it J. Light. Technol.}, {\bf 33,} 4377-4383 (2015).
	\bibitem{Akkaya}Akkaya, O. C., Akkaya, O., Digonnet, M. J., Kino, G. S., \& Solgaard, O. Modeling and demonstration of thermally stable high-sensitivity reproducible acoustic sensors. {\it J. Microelectromech. Syst.} {\bf 21,} 1347-1356 (2012).
	\bibitem{MaJun}Ma, J. et al. Fiber-optic Fabry–Perot acoustic sensor with multilayer graphene diaphragm. {\it IEEE Photon. Technol. Lett.} {\bf 25,} 932-935 (2013).
	\bibitem{Kilic}Kilic, O., Digonnet, M., Kino, G., \& Solgaard, O. External fibre Fabry–Perot acoustic sensor based on a photonic-crystal mirror. {\it Meas. Sci. Technol.} {\bf 18,} 3049 (2007).	
	\bibitem{Martin}Martin, D. T. et al. A micromachined dual-backplate capacitive microphone for aeroacoustic measurements. {\it J. Microelectromech. Syst.} {\bf 16,} 1289-1302 (2007).
	\bibitem{Bucaro}Bucaro, J. A., \& Lagakos, N. Lightweight fiber optic microphones and accelerometers. {\it Rev. Sci. Instrum.} {\bf 72,} 2816-2821 (2001).
	
	
	\bibitem{Preisser}Preisser, S. All-optical highly sensitive akinetic sensor for ultrasound detection and photoacoustic imaging. {\it Biomed. Opt. Express}, {\bf 7,} 4171-4186 (2016).
	\bibitem{QNoise}Gardiner, C. \& Zoller, P. {\it Quantum Noise: A Handbook of Markovian and Non-Markovian Quantum Stochastic Methods With Applications to Quantum Optics} (Springer Science \& Business Media, 2004).
	\bibitem{WallsMilburn}Walls, D.F. \& Milburn, G.J. {\it Quantum Optics} (Springer Science \& Business Media, 2007).
	\bibitem{BowenMilburn}Bowen, W.P. \& Milburn, G.J. {\it Quantum Optomechanics} (CRC Press, 2015).
	\bibitem{air-absorption}Bass, H. E., Sutherland, L. C., \& Zuckerwar, A. J. Atmospheric absorption of sound: Update. {\it J. Acoust. Soc. Am.}, {\bf 88,} 2019-2021 (1990).
	\bibitem{Bao}Bao, M. {\it Analysis and design principles of MEMS devices} (Elsevier, 2005).
	\bibitem{Kippenberg spoked toroids}Anetsberger, G., Rivi\`{e}re, R., Schliesser, A., Arcizet, O., \& Kippenberg, T. J. Ultralow-dissipation optomechanical resonators on a chip. {\it Nat. Photon.} {\bf 2,} 627 (2008).
	 \bibitem{Arcizet} Arcizet, O., Riviere, R., Schliesser, A., Anetsberger, G., \& Kippenberg, T. J. Cryogenic properties of optomechanical silica microcavities.   Physical Review A, {\bf 80}, 021803 (2009).
	\bibitem{Ho}Ho, Q. T., Verboven, P., Yin, X., Struik, P. C., \& Nicola\"\i, B. M. A microscale model for combined CO2 diffusion and photosynthesis in leaves. {\it PloS One}, {\bf 7,} e48376 (2012).
	\bibitem{Tam}Tam, A. C. Applications of photoacoustic sensing techniques. {\it Rev. Mod. Phys.} {\bf 58,} 381 (1986).
	\bibitem{Ahmed}Ahmed, A. et al. Biosensors for whole-cell bacterial detection. {\it Clin. Microbiol. Rev.
	} {\bf 27,} 631-646 (2014).
\bibitem{Longo}Longo, G. Rapid detection of bacterial resistance to antibiotics using AFM cantilevers as nanomechanical sensors. {\it Nat. Nanotechnol.} {\bf 8,} 522 (2013).
	\bibitem{Kasas}Kasas, S. et al. Detecting nanoscale vibrations as signature of life. {\it Proc. Natl. Acad. Sci. U.S.A.} {\bf 112,} 378-381 (2015).
	\bibitem{Lissandrello}Lissandrello, C. et al. Nanomechanical motion of Escherichia coli adhered to a surface. {\it Appl. Phys. Lett.} {\bf 105,} 113701 (2014).
	\bibitem{Song}Song, L. et al. Nanoscopic vibrations of bacteria with different cell-wall properties adhering to surfaces under flow and static conditions. ACS nano, {\bf 8,} 8457-8467 (2014).
\bibitem{Chan}Chan, J. et al. Laser cooling of a nanomechanical oscillator into its quantum ground state. {\it Nature}, {\bf 478,} 89 (2011).
	\bibitem{Teufel}Teufel, J. D. et al. Sideband cooling of micromechanical motion to the quantum ground state. {\it Nature}, {\bf 475,} 359 (2011).
	\bibitem{Schliesser}Schliesser, A. et al. Resolved-sideband cooling of a micromechanical oscillator. {\it Nat. Phys.} {\bf 4,} 415 (2008).
	\bibitem{Bowen}Lee, K. H., McRae, T. G., Harris, G. I., Knittel, J., \& Bowen, W. P.  Cooling and control of a cavity optoelectromechanical system. {\it Phys. Rev. Lett.} {\bf 104,} 123604 (2010).
\bibitem{Harris}Harris, G.I., Andersen, U.L., Knittel, J., \& Bowen, W. P.  Feedback-enhanced sensitivity in optomechanics: Surpassing the parametric instability barrier. {\it Phys. Rev. A} {\bf 85,} 061802(R) (2012).
\bibitem{Kim2017} Kim, P.H., Hauer, B.D., Clark, T.J., Fani Sani, F., Freeman, M.R., Davis, J.P. Magnetic actuation and feedback cooling of a cavity optomechanical torque sensor. {\it Nature Communications} {\bf 8} 1355 (2017).
\bibitem{Aspel} For a review see Aspelmeyer et al. Reviews of Modern Physics {\bf 86} 1391 (2014)

\end{references}
\end{document}